\documentclass{aastex61}

\usepackage{color}
\usepackage{amsmath}
\usepackage{array}
\newcolumntype{P}[1]{>{\centering\arraybackslash}p{#1}}

\usepackage[toc,page]{appendix}
\usepackage{afterpage}
\usepackage{float}
\usepackage{capt-of}
\usepackage{bm}

\received{August 8, 2017}
\accepted{May 11, 2018}
\shortauthors{Carr, Scarlata, Panagia \& Henry}
\begin{document}

\title{A Semi-Analytical Line Transfer (SALT) Model II: the effects of a Bi-Conical geometry}

\correspondingauthor{Claudia Scarlata}
\email{CodyCarr24@gmail.com, mscarlat@umn.edu}

\author{Cody Carr, Claudia Scarlata}
\affil{University of Minnesota,  316 Church str SE, Minneapolis, MN 55455,USA}

\author{Nino Panagia}
\affiliation{Space Telescope Science Institute, 3700 San Martin Drive, Baltimore, MD 21218, USA}
\affiliation{INAF--NA, Osservatorio Astronomico di Capodimonte, Salita Moiariello 16, 80131 Naples, Italy}
\affiliation{Supernova Ltd, OYV \#131, Northsound Rd., Virgin Gorda VG1150, Virgin Islands, UK}

\author{Alaina Henry}
\affiliation{Space Telescope Science Institute,3700 San Martin Drive,Baltimore, MD 21218, USA}



\begin{abstract}
We generalize the semi-analytical line transfer (SALT) model recently introduced by \citet{Scarlata:2015fea} for modeling galactic outflows, to account for bi-conical geometries of various opening angles and orientations with respect to the line-of-sight to the observer, as well as generalized velocity fields.  We model the absorption and emission component of the line profile resulting from resonant absorption in the bi-conical outflow. We show how the outflow geometry impacts the resulting line profile.  We use simulated spectra with different geometries and velocity fields to study how well the outflow parameters can be recovered.  We find that geometrical parameters (including  the opening angle and the orientation) are always well recovered. The density and velocity field parameters are reliably recovered when both an absorption and an emission component are visible in the spectra. This condition implies that the velocity and density fields for narrow cones oriented perpendicular to the line of sight  will remain unconstrained. 
\end{abstract}

\keywords{galaxies: intergalactic medium, galaxies: ISM, galaxies: starburst, ISM: jets and outflows}



\section{Introduction} \label{sec:intro}
As a prominent source of feedback, galactic winds play a vital role in the study of galaxy evolution and the enrichment of the intergalactic medium, IGM, \citep[][]{Veilleux:2005ia}.  Observations of low-ionization and high-ionization resonant transitions
in galactic winds often display P~Cygni type line profiles which can reveal a wealth of information regarding the physical nature of the winds.  Quantities of interest include density of the gas in the relevant ions, terminal velocity of the wind, and mass outflow rate.  

Galactic winds, in both the local universe \citep{shopbell1998asymmetric} and at higher redshift \citep{2012ApJ...760..127M,erb2015feedback}, appear to lack the symmetry of a full spherical outflow, and instead, are better described as having a collimated, or bi-conical geometry.  The geometry of the wind can have a strong impact on the estimation of galactic properties.  For example, \citet{2016arXiv160505769C} calculated up to an order of magnitude difference in their estimate of the mass loading factor (outflow rate divided by star formation rate) in NGC 6090 when comparing a spherical to a more realistic, bi-conical model. 

Various groups have modeled the absorption $+$ emission line profile resulting from the radiative transport of resonant photons in outflowing medium. Models range from simple analytic calculations \citep{martin2013scattered,Scarlata:2015fea,krumholz2017observable} to more advanced approaches involving Monte Carlo techniques \citep{2011ApJ...734...24P}.  This work is the second in a series of papers in which we explore how different geometries of the scattering medium affect the shape of the resulting line profile. In the first paper, we introduced the semi-analytical line transfer (SALT) model for resonant transitions observed in galactic spectra.  There, we explored the effects of having the density and velocity fields varying with radius, while still maintaining a simple spherical symmetry of the outflowing gas. In this second paper, we relax the spherical assumption, and model the line profiles resulting in bi-conical wind geometries, with diverse orientations with respect to an observer and opening angles. In both works we assume a constant ionization structure within the scattering wind, and -- as an example application-- we consider the resonant 1190\AA\ doublet of the Si$^{+}$ ion to illustrate the effects of the changing geometry on the line profiles. In followup papers, we will explore how the ionization structure of the wind  (which depends on the ionization source and the specific ion considered, i.e., ions with different ionization potentials behave differently) affects the density and velocity fields of the outflowing material, and use the models to reproduce actual galactic data.  In this paper we did not consider the effects of an infalling envelope on the absorption/emission line spectrum.  Infalling material would have an accelerated velocity field, with the velocity increasing as the gas gets closer to the galaxy.  In this situation, it can be shown that the gas is at resonance with a given transition in two regions, and a full radiative transport treatment would be required. 

This paper is organized as follows. We begin by reviewing the SALT model, originally presented by \citet{Scarlata:2015fea} for galactic outflows, in Section 2.  We generalize the SALT model to bi-conical outflow geometries characterized by an opening and an orientation angle in Section 3.  Our discussion follows in Section 4, where we investigate the impact of the outflow geometry on P~Cygni  line profiles. We test our ability to recover the wind parameters by fitting simulated data in Section 5.   Finally, we present our conclusions in Section 6.  

\section{Spherical SALT Model}
\begin{figure*}[t]
  \centering
\includegraphics[scale=0.4]{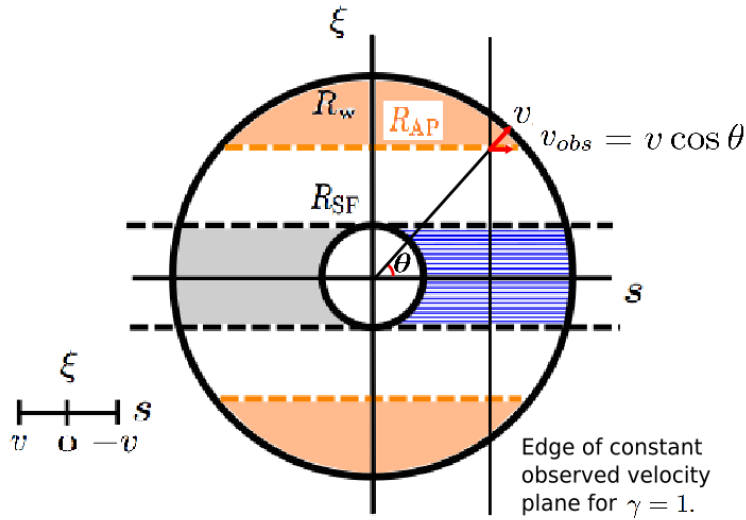}
 \caption{A cross sectional view of the spherical wind model.  The line of sight is indicated by the $s$-axis, and the $\xi$-axis is in the plane of the sky.  The small, central circle marks the star formation region/galaxy, with radius $R_{SF}$, while the outer circle indicates the extent of the wind ($R_{W}$).  The blue shaded region indicates the absorption region. 
 The orange areas indicate those parts of the outflowing envelope blocked by a circular spectroscopic aperture of radius $R_{AP}$.   An occulted region is shaded in grey, since emission cannot be detected from behind the galaxy.   A velocity vector, $v$, along with its component along the line of sight, $v_{\text{obs}}$, is shown in red.  The thin vertical line represents the edge of the plane of constant $v_{obs}$ for a $\gamma =1$ velocity field. }
   \label{fig:sphere_crosssec2.png}
\end{figure*}

In this section, we review the SALT model introduced by \citet{Scarlata:2015fea} for spherical galactic outflows.  The fiducial model consists of a spherical source (radius $R=R_{SF}$) of isotropically emitted radiation.  Physically, the source embodies the star formation region/galaxy.   The actual geometry of the source has a negligible effect on the outcome of this work, however, it is important to realize that it is an artificial construct.  When the star formation region is clearly not a sphere (e.g. in thin-disk galaxies), $R_{SF}$ is better understood as an average of the disk radii projected onto the plane of the sky.  The calculations in this work will be extended to account for non spherical geometries of the emitting continuum source in a future paper.

The source is located at the center of a spherical envelope (wind) of material extending from the radius $R_{SF}$ to the terminal radius, $R_W$.  A schematic representation of a cross sectional cut of the envelope is presented in Figure \ref{fig:sphere_crosssec2.png}.   The $\xi$-axis runs perpendicular to the line of sight and is measured using normalized units, $r/R_{SF}$.  The $s$-axis runs parallel with the line of sight and is measured using the same normalized units.  We refer to an arbitrary radius measured in the normalized units as $\rho$.     

The wind is characterized by a velocity field, $v$, and a density field, $n$.  We assume a power law for the velocity field, with:  
\begin{equation}
\begin{aligned}
v &= v_0\left(\frac{r}{R_{\text{SF}}}\right)^{\gamma} &&\text{for}\ r < R_{W} \\[1em]
v &= v_{\infty}  &&\text{for} \ r \geq R_{W}, \\
\end{aligned}
\end{equation}

\noindent  where $v_0$ is the wind velocity at the surface of the source (i.e., at $R_{\text{SF}}$), and $v_{\infty}$ is the terminal velocity of the wind at $R_{W}$.  Assuming a steady outflow rate, the density and velocity are related by: $n \propto (vr^2)^{-1}$.  We assume that throughout the outflow the ionization structure of the gas remains constant. A more detailed exploration of how the ionization structure (which depends on the specific ionizing source as well as on the elements considered) affects the velocity and density fields is beyond the scope of the current work, and will be presented in a forthcoming paper. 

To solve the radiation transfer problem, we adopt the Sobolev's approximation \citep{ambartsumian1958theoretical,sobolev1960moving}. Provided the velocity gradient, $dv/dr$, is large, an atom moving with respect to the source will only absorb a photon at resonance with a transition line.  Mathematically, we can approximate the profile function of the optical depth as a delta function evaluated at the Doppler shifted resonance frequency \citep{1999isw..book.....L}.  

In essence, we have reduced the radiation transfer problem to a local problem and can now decompose the outflow into thin spherical shells of a given radius, $r$, velocity, $v$, and optical depth, $\tau(r,\phi)$.  Here, $\phi$ is the angle between the velocity and the trajectory of the photon.  The optical depth is 

\begin{eqnarray}
\tau(r) &=& \frac{\pi e^{2}}{mc}f_{lu}\lambda_{lu}n_{l}(r)\left[1 - \frac{n_{u}g_{l}}{n_{l}g_{u}}\right]\frac{r/v}{1 + \sigma \cos^2{\phi}},
\end{eqnarray}
\noindent where $f_{ul}$ and $\lambda_{ul}$ are the osciallator strength and wavelength, respectively, for the $ul$ transition, and $\sigma = \frac{\text{d}ln(v)}{\text{d}ln(r)}-1$  \citep{castor1970spectral}.  Assuming our density and velocity field hold, and by neglecting stimulated emission (i.e., $\left[1 - \frac{n_{u}g_{l}}{n_{l}g_{u}}\right] =1$), we get 

\begin{eqnarray}
\tau(r)&=& \frac{\pi e^{2}}{mc}f_{lu}\lambda_{lu}n_{0}\left(\frac{R_{\text{SF}}}{r}\right)^{\gamma+2}\frac{r/v}{1 + \sigma  \cos^2{\phi}}\\
&=& \frac{\tau_{0}}{1+(\gamma-1)\cos^2{\phi}}\left(\frac{R_{\text{SF}}}{r}\right)^{2\gamma+1}
\end{eqnarray}
\noindent and
\begin{eqnarray}
\tau_{0} = \frac{\pi e^{2}}{mc}f_{lu}\lambda_{lu}n_{0}\frac{R_{\text{SF}}}{v_{0}},
\end{eqnarray}
where $ n_{l}(r) = n_0\left(\frac{R_{SF}}{r}\right)^{\gamma+2}$ and $n_0$ is the gas density at $R_{SF}$.  By allowing $\gamma$ to vary, we will need to make a few slight modifications to several expressions presented in \citet{Scarlata:2015fea}, who focus on the $\gamma = 1$ scenario.  For this reason, we have placed a brief review of their procedure for multiple scattering of reemitted photons in the appendix along with our modifications.         

The observed spectrum needs to be computed in terms of  observed velocities, $v_{\text{obs}}$, which, for a given shell, is the projection of its intrinsic velocity onto the line of sight (i.e., $v_{\text{obs}} = v \cos \theta$, where $\theta$ is the angle between the velocity at a given position and the line of sight). At each $v_{obs}$, the observed spectrum is then computed accounting for all shells that absorb and reemit at that observed velocity.  For a single shell, regions of constant observed velocity form rings in the plane of the sky.  Within the outflow, surfaces of constant $v_{obs}$ can be formed by connecting all rings of the same observed velocity from neighboring shells.  

An infinitesimal shell will absorb a fraction $E(\tau)=1-e^{-\tau(v)}$ of the emitted energy. For a spherical shell, this absorbed energy will be evenly distributed in terms of the observed velocities.  (We will revisit this statement later when we generalize to non-spherical outflow geometries.) 

$\bullet$ {\bf Absorption component} To compute the absorption component of the observed spectrum, we need only concern ourselves with the portion of each shell in front of the source, as viewed on the plane of the sky (i.e., in the hatched region). A given shell can only absorb in the range of observed velocities between $[v, v_{\text min}]$, where $v_{\text{min}}$ is the component of the velocity along the line of sight computed at $\xi=1$.
Setting $y = v/v_0$, we compute:

\begin{eqnarray}
y_{\text{min}} &=&\frac{v_{min}}{v_0}= y \cos{\theta} \\
&=& y^{(\gamma - 1)/\gamma}(y^{2/\gamma}-1)^{1/2}.
\end{eqnarray}

\noindent  
All shells with intrinsic velocity between $v_{\text{obs}}$ and $v_1=v_{\text{obs}}/\cos{\theta}$ contribute to the absorption at $v_{obs}$. Setting $x = v_{\text{obs}}/v_0$ and $y_1 = v_1/v_0$ , one can solve the following equation to get $y_1$:
\begin{eqnarray}
y_{1}^{2}(1-y_{1}^{-2/\gamma}) = x^2.
\end{eqnarray}

Thus, the normalized absorption profile becomes 
\begin{equation}
I(x)_{\rm{abs, blue}} = 1 - \int_{\rm{max}(x,1)}^{y_1}\frac{1-e^{-\tau(y)}}{y- y_{\rm{min}}}dy,
\end{equation}
\noindent where the lower bound of integration excludes the source.  

$\bullet$ {\bf Emission component}. We have separated the emission profile into red shifted (positive velocity) and blue shifted (negative velocity) components.  For both profiles, half of the remitted energy will be spread evenly from 0 to y.  For a given $v_{obs}$ in the red component, we exclude contributions from all shells where the location of $v_{obs}$ is blocked from the observer's field of view by the source. Thus, only shells with intrinsic velocities from $y_1$ to $y_{\infty} = v_{\infty}/v_0$ will contribute to a given $v_{obs}$.  The blue component will not be occulted, hence, all shells, excluding the source, will contribute. The resulting normalized profiles for the red and blue components are               
\begin{equation}
I(x)_{\rm{em, red}} = \int_{y_1}^{y_{\infty}}\frac{1-e^{-\tau(y)}}{2y}dy,
\end{equation}
\noindent
and 
\begin{equation}
I(x)_{\rm{em, blue}} = \int_{\rm{max}(x,1)}^{y_{\infty}}\frac{1-e^{-\tau(y)}}{2y}dy,
\end{equation}
\noindent respectively.  The full P~Cygni profile becomes
\begin{eqnarray}
I(x) =   I(x)_{\rm{abs, blue}}+ I(x)_{\rm{em, red}}+I(x)_{\rm{em, blue}}.
\end{eqnarray}

In this formalism it is easy to account for an envelope that does not cover the full source.  We define $f_c$ as the wind covering fraction.  We keep $f_c$ constant for all shells.  $f_c$ can be thought of as clumps or small holes in the wind uniformly distributed across each shell -- although the global fraction of energy absorbed by a shell is now $f_cE(\tau)$, the energy will still be evenly distributed.  

Lastly, we need to account for a limiting, circular observing aperture of radius, $R_{AP}$.  Intuitively, one would expect the emission component of the spectrum to be underestimated when the aperture size is smaller than the wind.  One can account for this by changing the range of integration for each profile \citep{Scarlata:2015fea}, however, the same effect can be accounted for by removing shells from the outflow using a scaling factor.  We define the aperture factor
\begin{eqnarray}
\Theta_{AP} &:=& \Theta(y_{\text{ap}} - [y^{2/\gamma}-x^2y^{2(1-\gamma)/\gamma}]^{\gamma/2} ),
\end{eqnarray}

\noindent
where $\Theta$ is the Heaviside function:

\begin{gather} 
\Theta := 
\begin{cases}
0 & \rm{if \ }  \ y_{\text{ap}}  < [y^{2/\gamma}-x^2y^{2(1-\gamma)/\gamma}]^{\gamma/2} \\[1em]
1 & \rm otherwise.\\
\end{cases}
\end{gather}
$y_{ap} = v_{ap}/v_0$, where $v_{ap}$ is the intrinsic velocity of the shell with radius $R_{AP}$.  This scale factor will remove all shells that fall outside the aperture radius for a given, $x$.  For an explicit derivation of $\Theta_{AP}$, we refer the reader to the appendix.  

Finally, combining the wind covering fraction, $f_c$, and the aperture factor, $\Theta_{AP}$, the normalized spherical (Sp) profiles are:

\begin{equation}\label{3}
\begin{aligned}
I(x)_{\rm{Sp}} &=& 1 - \int_{\rm{max}(x,1)}^{y_1}\frac{\Theta_{AP}f_c(1-e^{-\tau(y)})}{y- y_{\rm{min}}}dy\\[1em]
&+& \int_{y_1}^{y_{\infty}}\frac{\Theta_{AP}f_c(1-e^{-\tau(y)})}{2y}dy\\[1em]
&+& \int_{\rm{max}(x,1)}^{y_{\infty}}\frac{\Theta_{AP}f_c(1-e^{-\tau(y)})}{2y}dy.\\
\end{aligned}
\end{equation} 

\section{Bi-conical SALT Model}

\begin{figure*}[t]
  \centering
\includegraphics[scale=0.7]{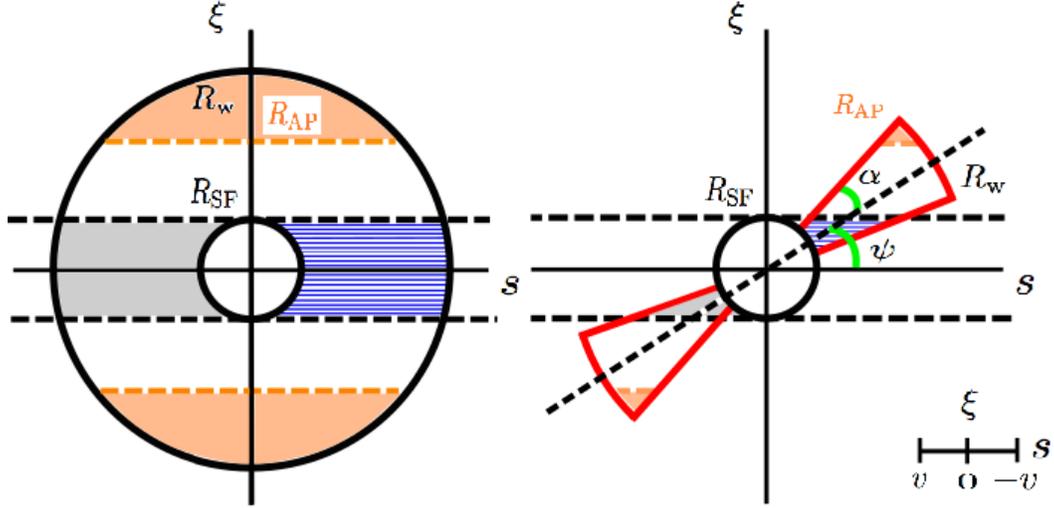}
 \caption{$Left$, same as Figure~1. $Right$, cross sections for a bi-conical outflow.  The different colored shades define the same regions as described in Figure~\ref{fig:sphere_crosssec2.png}.  The wind extends from the radius $R_{SF}$ to the terminal radius $R_{W}$.  The bi-conical outflow, shown in red, is defined by the orientation angle, $\psi$, and the opening angle, $\alpha$.} 
   \label{fig:f_oaprofile}
\end{figure*}

	We have constructed a bi-conical outflow with the cones' focal point positioned at the center of the source in Figure \ref{fig:f_oaprofile}.  The cones are described by two parameters: $\alpha$ and $\psi$.  $\alpha$ is the half opening angle for each cone.  We will refer to $\alpha$ as the opening angle.  $\psi$ is the overall angular displacement between the line of sight and the axis of the cones.  We will refer to $\psi$ as the orientation angle.  A single shell of velocity $v$ now consists of the portion of a spherical shell that lies within the bi-conical outflow.  We will refer to it as a bi-conical shell.  A blue hatched section has been drawn to indicate the new absorption region.  

	To account for a bi-conical geometry, we must consider how the energy absorbed by a given shell will be distributed in terms of the observed velocities. For a spherical shell of  velocity $v$ the energy is distributed uniformly between $\pm v$. This is demonstrated below and schematically shown in the top left panel of  Figure~\ref{fig:constant_vel}, which shows the emission line profile for a thin spherical shell. Depending on the orientation and opening angle, a bi-conical shell with velocity $v$ may distribute the absorbed energy over a smaller observed velocity range. In the remaining panels of Figure~\ref{fig:constant_vel}, we show how the line profile changes for three bi-conical shells all with the same velocity $v$, but different orientations and opening angles.  In addition to not covering the full $\pm v$ velocity range, the energy is no longer evenly distributed.  

To see where the latter effect comes from, we first show why the line profile of a spherical thin shell is flat between $\pm\, v$.
	
\begin{figure*}[t!]
  \centering
\includegraphics[scale=0.55]{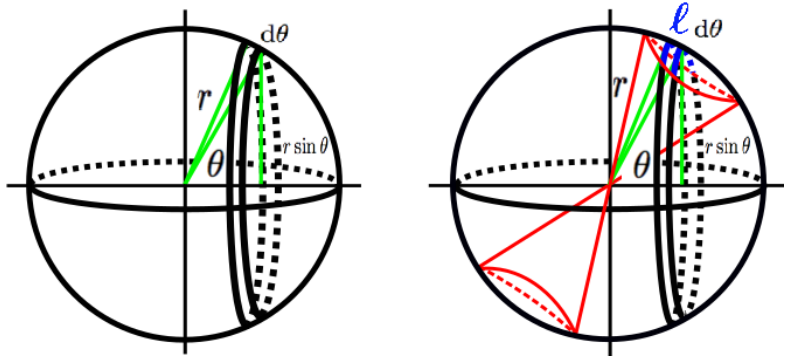}
 \caption{Band contours of constant observed velocity are shown for spherical (left) and bi-conical (right) outflows.  The geometric factor, $f_g$, is defined as the fraction of the spherical  band contours subtended by the bi-conical shell.  The overlap region is shown in blue.}
 \label{fig:contours}
\end{figure*}

	Following \citet{1931MNRAS..91..966B}, we note that the energy absorbed by a band contour, or ring of constant observed velocity (see Figure \ref{fig:contours}), will be proportional to the area of the band, i.e., 
\begin{eqnarray}
\text{d}I(v_{\text{obs}})_{\text{Sp}}&=& (\text{constant}) r\text{d}\theta 2 \pi r \sin{\theta}.
\end{eqnarray}
\noindent Differentiating the observed velocity, $v_{\text{obs}} = -v\cos{\theta}$, we get

\begin{eqnarray}
\text{d}v_{\text{obs}}&=& v \sin{\theta} \text{d} \theta.
\end{eqnarray}
{\bf Hence,}
\begin{eqnarray}
\frac{\text{d}I(v_{\text{obs}})}{\text{d}v_{\text{obs}}}_{\text{Sp}} &=& (\text{constant})2\pi\frac{r^2}{v},
\end{eqnarray}
\noindent {\bf and the distribution is independent of $\theta$.  Thus, the energy absorbed by a single shell of velocity v and radius r will be evenly distributed in terms of the observed velocities.} 

\begin{center}
\includegraphics[scale=.45]{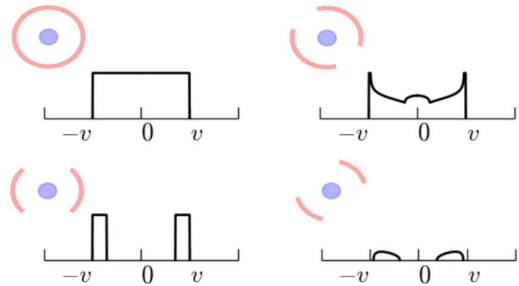}
\captionof{figure}{Examples of line profiles for the emission component (in observed velocity space) produced by thin shells (all with the same intrinsic velocity, $v$), for various outflow geometries (neglecting occultation from the source).  The upper left plot shows the emission profile produced by a spherical thin shell, with the constant emissivity as a function of $v$, extending from $\pm v$. In the remaining three plots we consider bi-conical shells with different orientations with respect to the line of sight and different geometries.  Bottom left: emission from a thin shell in a bi-conical outflow oriented along the line of sight.  The emission profile is constant in the velocity range between $v$, and the observed velocity corresponding to the edge of the shell (which does not extend all the way to $0$). Bottom right: same as before, but now the outflow is oriented $45^{\circ}$ away from the line of sight. The emission profile is modulated by the projection of the shell on the line of sight, it never reaches either zero or $v$. Top right: a bi-conical shell oriented $45^{\circ}$ away from the line of sight, but with a larger opening angle than before.  At $v_{obs}\sim v$, the emission is the same as that of a full sphere.}
 \label{fig:constant_vel}
\end{center}

	In the case of a bi-conical shell the band contour will no longer be circular, i.e., at a given observed velocity, it will have an arc length, $l < 2\pi r \sin{\theta}$. Hence, with BC denoting the  bi-conical case:
\begin{eqnarray}
\text{d}I(v_{\text{obs}})_{\text{BC} }&=& \text{(constant)}r\text{d}\theta l  \\
\implies \text{d}I(v_{\text{obs}})_{\text{BC} } &=& \frac{l}{2\pi r\sin{\theta}} \text{d}I(v_{\text{obs}})_{\text{Sp}}\\
\implies \frac{\text{d}I(v_{\text{obs}})}{\text{d}v_{\text{obs}}}_{\text{BC} }  &=&f_g\frac{\text{d}I(v_{\text{obs}})}{\text{d}v_{\text{obs}}}_{\text{Sp}},
\end{eqnarray}
\noindent where $f_g = l/2\pi r \sin{\theta}$.  Thus, the bi-conical energy distribution can be obtained by scaling the spherical energy distribution by a scale factor, $f_g$.  We call this scaling ratio the geometric factor.  Note that the arc length, $l$, is a non-trivial function of the observed velocity, $x$, implying that the absorbed energy will no longer be evenly distributed in terms of $x$ as in the spherical case.  We have provided an explicit calculation of $f_g$ in the appendix.

To conclude, the normalized bi-conical profiles are:

\begin{equation}\label{4}
\begin{aligned}
I(x)_{\rm{BC}} &=& 1 - \int_{\rm{max}(x,1)}^{y_1}\frac{\Theta_{AP}f_g(1-e^{-\tau(y)})}{y- y_{\rm{min}}}dy\\[1em]
&+& \int_{y_1}^{y_{\infty}}\frac{\Theta_{AP}f_g(1-e^{-\tau(y)})}{2y}dy\\[1em]
&+& \int_{\rm{max}(x,1)}^{y_{\infty}}\frac{\Theta_{AP}f_g(1-e^{-\tau(y)})}{2y}dy.\\
\end{aligned}
\end{equation}

\section{Discussion}

\subsection{Line Profiles for Bi-conical Outflows}

%	In this section, we analyze the two models presented in equations \ref{3} and \ref{4} of the previous section.  

\begin{figure*}[tp]
  \centering
%\begin{minipage}[t]{0.45\linewidth}
\includegraphics[scale=0.3]{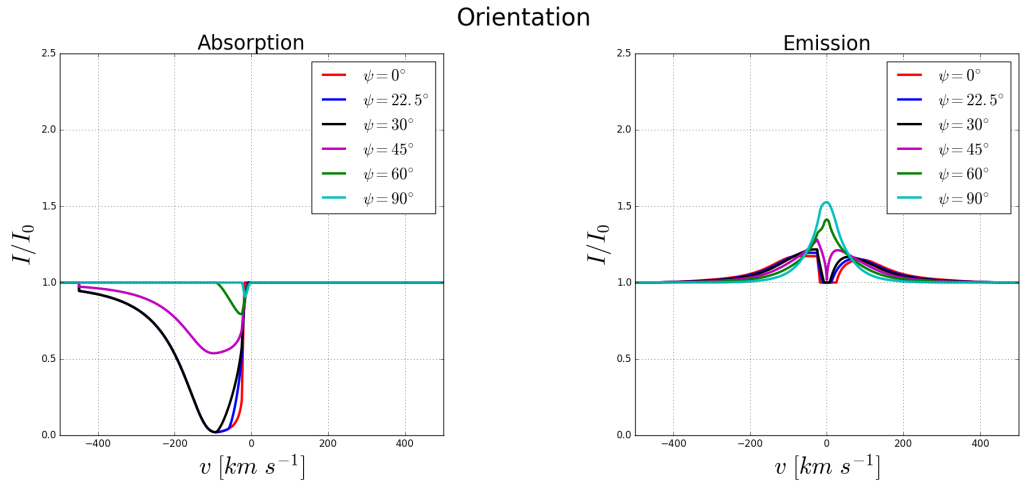}
 \caption{Absorption (left) and emission (right) component of the profiles generated in a bi-conical outflow model, as the orientation of the cone with respect to the line of sight is varied, while all other parameters are kept fixed ($\tau_0 = 330, \ \gamma = 1, \  v_0 = 25 \ km \ s^{-1}, \ v_{\infty}=450 \ km \ s^{-1}$, and $\alpha=45^{\circ}$).  The orientation angle, $\psi$, ranges from $0^{\circ}$, along the line of sight, to $90^{\circ}$, perpendicular to the line of sight.  As $\psi$ increases, the outflow moves away from the line of sight, resulting in fewer and fewer shells contributing to the absorption (left), and less emission at high projected velocities (right).  The emission profiles are double peaked because no material is reemitting at projected velocities of $zero$ for small inclination angles. At the limit of $\psi =90^{\circ}$, the emission component is a single peak, centered at zero. \\} 
   \label{fig:abspsi1}
%\end{minipage}

 \end{figure*}

	We begin the discussion by investigating how the geometry of the outflow (i.e., the opening angle $\alpha$ and orientation angle $\psi$) impacts the shape of the resonant line profiles.  In Figure~\ref{fig:abspsi1}, we vary the orientation angle while the opening angle remains fixed at $45^{\circ}$.  We show the absorption and emission components of the profiles separately in the left and right panels of the Figure, respectively.  As $\psi$ increases, the outflow covers a smaller fraction of the source, resulting in a smaller overall absorption. For large $\psi$, the absorption is limited to small observed velocities, reflecting the fact that the outermost shells (i.e., with the higher velocities) are not contributing to the absorption. For $\psi=90^{\circ}$, the outflow is oriented perpendicularly to the line of sight, and the absorption is almost negligible. At this orientation the emission component is symmetric with respect to the systemic velocity (if we ignore the small occulting effect of the source). As $\psi$ decreases and the outflow rotates towards the line of sight,  gas is no longer moving perpendicularly to it. Therefore the reemission at zero velocity decreases, and a dip appears in the emission profile. The dip broadens and the emission at larger observed velocities increases as the outflow continues to move closer to the line of sight.  Finally, it is important to realize that for specific inclination/opening angle combinations of the bi-conical outflow, the maximum  velocity at which absorption occurs may largely underestimate  the true terminal velocity of the wind.  This happens when the forward facing part of the outflow is oriented in such a way that absorption  no longer occurs at the terminal velocity (e.g., when $\psi > \alpha$, if $y_{\infty}^{1/\gamma} \sin{(\psi - \alpha )} > (\xi = 1)$). For these orientations, $v_{\text{max}}$ only reflects the outflow geometry and orientation.  
	
\begin{figure*}[tp!]
  \centering

\includegraphics[scale=0.3]{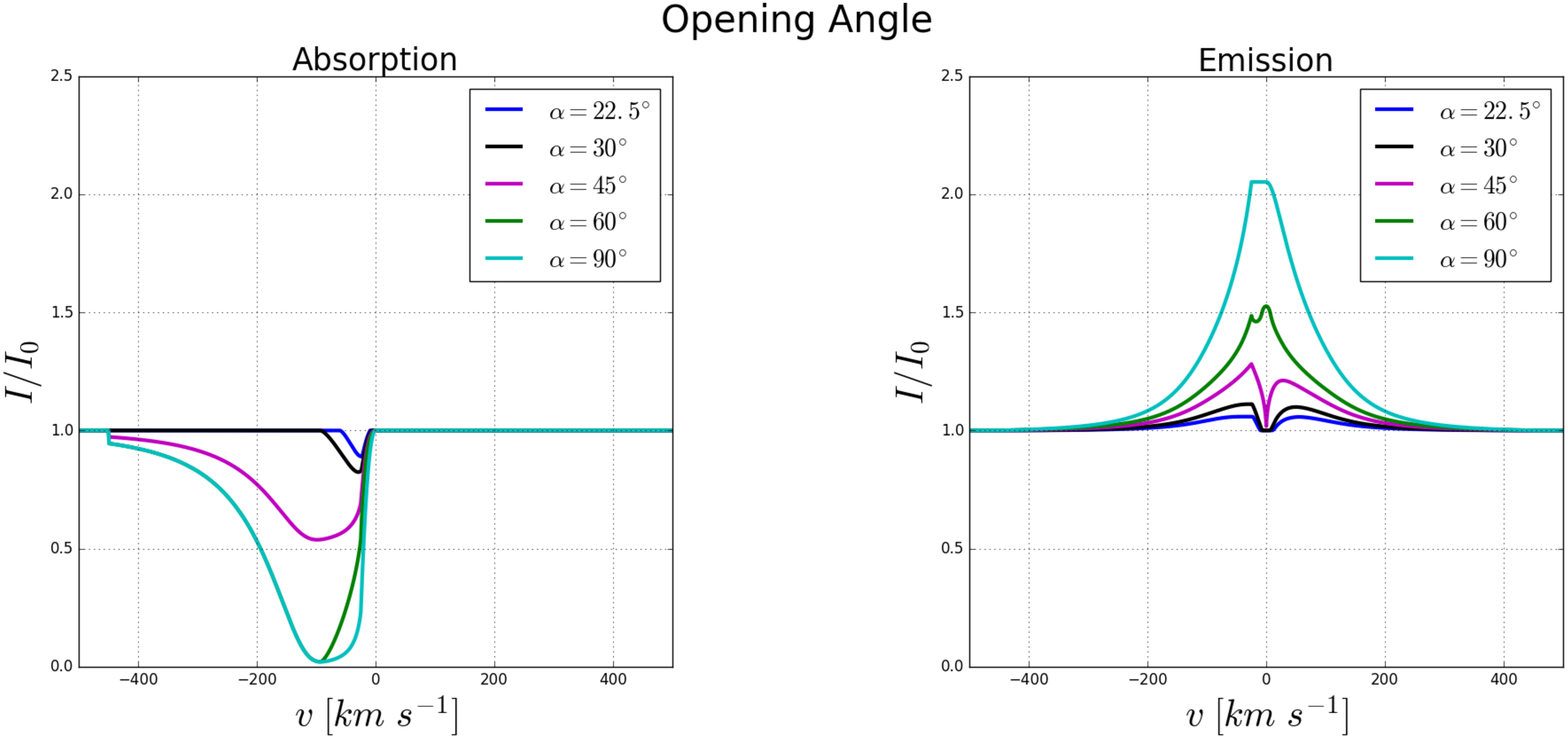}
 \caption{Same as Figure~\ref{fig:abspsi1}, but now varying the cone opening angle, $\alpha$, in the range $22.5^{\circ}$ to $90^{\circ}$ (which returns the spherical model). As $\alpha$ decreases, shells will cover a smaller fraction of the source, resulting in weaker absorption (left).  For small $\alpha$,  a dip in emission (right) occurs near zero velocity when $\alpha$ is small enough such that the outflow is no longer perpendicular to the line of sight.  (The remaining parameters used to generate the profiles are: $\tau_0 = 330, \ \gamma = 1, \ v_0 = 25  \ km \ s^{-1}, \ \rm{and} \ v_{\infty}= 450 \ km \ s^{-1}$.) } 
   \label{fig:empsi1}
\end{figure*}

	We consider the opposite scenario in Figure~\ref{fig:empsi1}, where we vary the opening angle and keep the orientation angle fixed at $45^{\circ}$.  For $\alpha  = 90^{\circ}$, the geometry is spherical and the profiles are the same as presented in \citet{Scarlata:2015fea}.  As $\alpha$ decreases, the outflow decreases in overall size, covering a smaller fraction of the source, leading to a smaller overall absorption component.  For small $\alpha$, absorption occurs only at lower velocities because the shells at larger velocity are no longer in front of the source.  In the right panel, we see that for $\alpha < 90^{\circ} - \psi$, material no longer moves perpendicular to the line of sight, resulting in an overall decrease in emission at zero observed velocity.  The emission dip will broaden as $\alpha$ continues to decrease as fewer shells contribute at low velocities.  Similar to Figure~\ref{fig:abspsi1}, the asymmetry between the blue and red emission components is due to occultation by the source.     
	
\begin{figure*}[tp!]
  \centering
\includegraphics[scale=0.3]{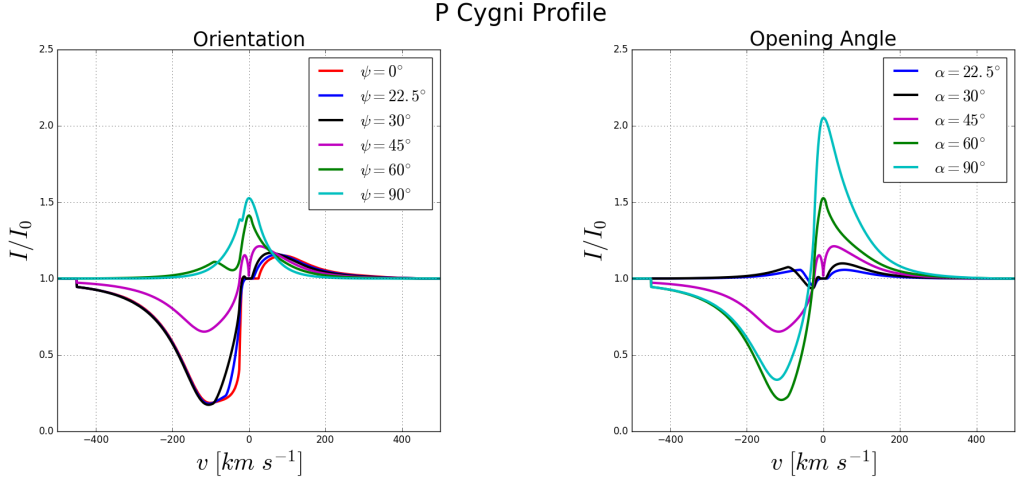}
 \caption{Full resonant line profile (i.e., absorption $+$ emission) for variable orientation angle (left) and variable opening angle (right).  } 
   \label{fig:total}
%\end{minipage}
\end{figure*}

	  The full P~Cygni profile (i.e., absorption $+$ emission) for variable orientation angle and variable opening angle are shown in the left and right panels of Figure~\ref{fig:total}, respectively.  These figures show that the apparent absorption component is decreased as a result of the infilling by the blue emission.  However, many of the same features from the previous discussion are still visible including the emission dip.  Observing such a feature would be strong evidence for a bi-conical outflow \citep{bae2016prevalence}.
	  
\begin{figure*}[tp!]
  \centering
   \includegraphics[scale=0.3]{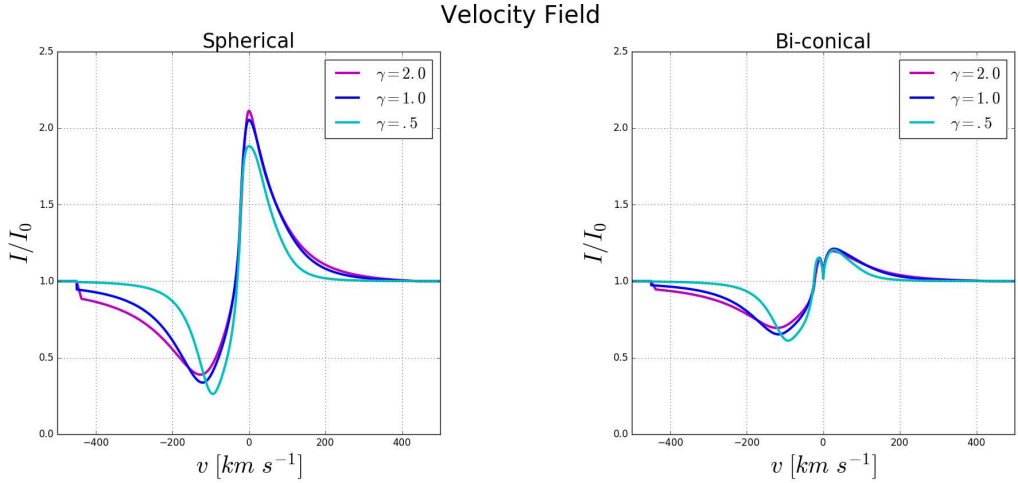}
 \caption{Full resonant line profiles computed varying the power law index ($\gamma$) of the velocity field.  Spherical and bi-conical outflows are shown on the left and right panel, respectively.  The bi-conical outflow used has geometry $(\alpha,\psi) = (45^{\circ},45^{\circ})$, while the remaining parameters used to generate the profiles are: $\tau_0 = 330, \ v_0 = 25 \ km \ s^{-1}, \ \rm{and}  \ v_{\infty}=450 \ km \ s^{-1}$. As $\gamma$ increases, the terminal velocities are reached at progressively smaller distances and progressively larger densities, which results in stronger absorption for larger $\gamma$.  For the bi-conical outflow, both how quickly the density grows with respect to the velocity field and the geometry will affect how much absorption/re-emission occurs at a given observed velocity. } 
\label{fig:velocity_field_spherical}
\end{figure*}

	  	We consider the effects of varying the velocity power law index, $\gamma$, in Figure~\ref{fig:velocity_field_spherical}, where we consider the spherical and bi-conical outflows in the left and right panels, respectively.  In calculating the profiles we only change the value of $\gamma$. In both the spherical and bi-conical cases,  the strength of the absorption at high observed velocities increases with $\gamma$. This is because, as $\gamma$ increases, the terminal velocity (which is constant in all models) is reached at progressively smaller distances, i.e., at progressively larger densities -- resulting in  stronger absorption for larger $\gamma$. Consequently, the stronger absorption is reflected in a stronger re-emission component. In the bi-conical geometry two effects are responsible for the observed line profile: 1) how quickly the density grows with respect to the velocity field, and 2) how much of the wind contributes to the emission at a given observed velocity.   The first effect is controlled by $\gamma$ and $\tau$.  The second effect is controlled by the geometry/orientation of the outflow and $\gamma$ -- which define the surfaces of constant observed velocity (see Figure~\ref{fig:concentric_shells} in the Appendix). 
			
\begin{figure*}[tp!]
  \centering
\includegraphics[scale=0.3]{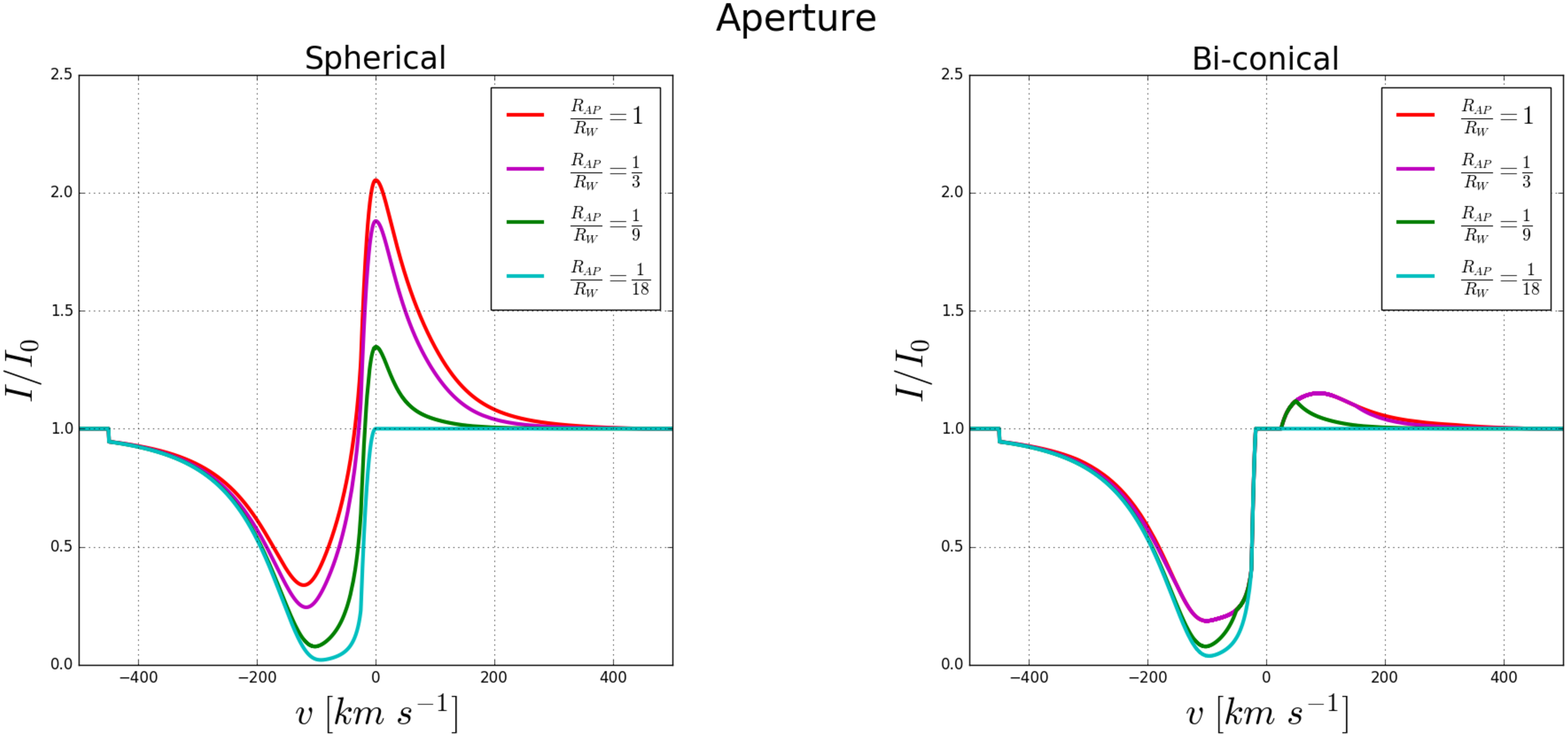}
 \caption{Full resonant line profiles computed varying the size of the spectroscopic aperture ($R_{AP}$). Spherical and bi-conical outflows are shown on the left and right panel, respectively.  The bi-conical outflow used has geometry $(\alpha,\psi) = (45^{\circ},0^{\circ})$, while the remaining parameters are the same as those used in Figure~\ref{fig:velocity_field_spherical}. The ratio $R_{AP}/R_{W}$ represents the fraction of the outflow  within the observing circular aperture centered on the source.  At $R_{AP}/R_{W} = \frac{1}{18}$, the aperture is the same size as the source, i.e., $R_{AP} = R_{SF}$.  As $R_{AP}$ decreases, the emission profile diminishes rapidly.  In the bi-conical outflow, as $R_{AP}$ decreases, the emission profile begins to decrease, however, the effect is less prominent compared to the spherical model. The majority of the wind lies along the line of sight and is not eminently affected by the decreasing aperture.  } 
   \label{fig:aperture}
\end{figure*}

\subsection{Effects of  Aperture and  Covering Fraction}
Spectroscopic observations are typically conducted with an aperture of finite size, which may cover only part of the scattering envelope. The shape of the resulting line profile will depend on the shape of the aperture, and on the relative sizes of the aperture and the envelope. We explore here the effects of circular apertures. We show the results for  spherical and bi-conical outflows in Figure~\ref{fig:aperture}, left and right panels, respectively.  The opening angle of the bi-conical outflow is fixed at $45^{\circ}$ and the outflow is oriented parallel to the line of sight.  As expected, for apertures smaller than the terminal radius of the wind, i.e., for $R_{AP}<R_{W}$,  the overall emission decreases, for both the spherical and bi-conical models.  
The shape of the profile also changes, i.e., the decrease in emission is not a simple scaling factor. This velocity dependence is introduced by the fact that smaller apertures progressively block photons scattered by regions with different projected velocities. This effect can be clearly seen  in the bi-conical case example shown in Figure~\ref{fig:aperture}, where photons emitted by regions at progressively smaller projected velocities are blocked. Smaller apertures also help to reduce the contribution of the blue component of the scattered re-emission to the absorption profile \citep[effect known as "in-filling", e.g., ][]{2011ApJ...734...24P,2012ApJ...760..127M,Erb:2012tp}.

\begin{figure}[tp!]
  \centering
\includegraphics[scale=0.3]{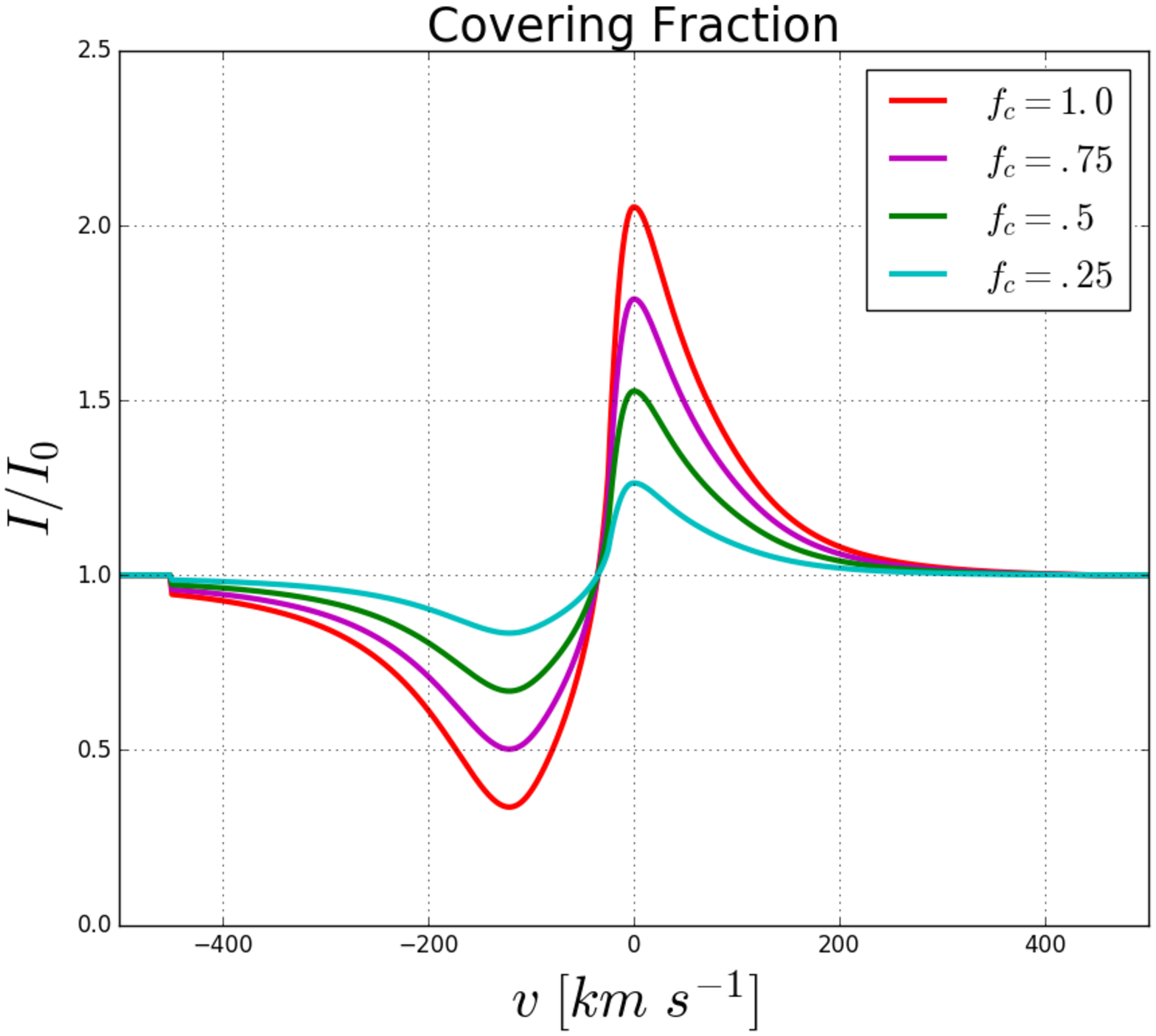}
 \caption{Full resonant line profiles resulting in a spherical outflow, computed varying the outflow covering fraction, $f_c$.  For $f_c = 1$, there are no holes/clumps in the wind and we recover the fiducial model.  Reducing $f_c$ diminishes both the emission and absorption components.  This is distinguishable from the effect of reducing the aperture, which reduces only emission. (The remaining parameters used to generate the profiles are: $\tau_0 = 330, \ \gamma = 1, \  v_0 = 25 \ km \ s^{-1}, \ \rm{and}  \ v_{\infty}=450 \ km \ s^{-1}$.) \\} 
   \label{fig:ccf}
     \end{figure} 
%     
%\begin{figure*}[tp!]
%  \centering
%\includegraphics[scale=0.3]{Sphere_fc_tau.png}
% \caption{Full resonant line profiles resulting in a spherical outflow, computed varying the outflow covering fraction, ($f_c$, left) and optical depth (right, for $f_c = 0.5$).  For $f_c = 1$, there are no holes/clumps in the wind and we recover the fiducial model.  Reducing $f_c$ diminishes both the emission and absorption components.  This is distinguishable from the effect of reducing the aperture, which reduces only emission.  In the right panel we fix $f_c = .5$, and allow the optical depth, $\tau$, to vary as multiples of $\tau_0$.  Increasing $\tau$ does lead to saturation in absorption like one usually expects, however, $f_c$ will always leave a portion of the source uncovered, resulting in absorption components which do not reach zero.  (The remaining parameters used to generate the profiles are: $\tau_0 = 330, \gamma = 1,  v_0 = 25 \rm{ \ km/s}, v_{\infty}=450 \rm{ \ km/s}$.) \\} 
%   \label{fig:ccf}
%     \end{figure*} 
%     
Recent works have advocated a covering fraction ($f_c$) smaller than unity for the neutral outflowing gas  \citep{rivera2015lyman,jones2013keck}. We have explored this possibility in Section~2.1, where we computed the profiles generated in a spherical outflow with varying $f_c$. The resulting line profiles are shown in Figure~\ref{fig:ccf}, left panel. The effect of a decreasing covering fraction is to reduce both the absorption and the scattered re-emission components.  This effect is distinguishable from the changes introduced by different sized apertures, which act only to reduce the emission component (leaving the absorption untouched).  A bi-conical outflow effectively covers only a fraction of the emitting source.  The profiles resulting in the two cases (spherical with $f_c \neq 1$ and bi-conical), however, can clearly be distinguished, by the shape of the emission profile (e.g., the dip in the emission component at zero projected velocities, see Figure~\ref{fig:total}) and the ratio between the areas of the absorption and emission components. 
%
%In the right panel of Figure~\ref{fig:ccf}, we fixed the covering fraction to $f_c = .5$ and explore how changing the optical depth, $\tau_0$, changes the shape of the resulting profile.  As the absorption saturates, i.e., $e^{-\tau} \rightarrow 0$ in $I/I_{0}$, the fraction of energy absorbed by a single shell $E(v) \rightarrow f_c$.   The maximum possible depth of the absorption dip becomes $1-f_c$ because a portion of the source always remains uncovered, allowing radiation to pass undisturbed directly to the observer. 	
		
\subsection{Equivalent Widths and Blue Emission In-fill}

	We next quantify the effects of the geometry on the ratio between the absorption and emission equivalent widths (EWs) of the model profiles. To do so,  we directly compare the area produced by the light profile below the continuum, $A_{ABS}$, to the total area above the continuum, $A_{EM}$ (which includes both the  red and blue emission components).   $A_{ABS}$ is defined to be negative. To study the effect of the emission on the absorption profile as a function of the outflow geometry ($\alpha,\psi$), we compute the ratio $R_{EW}$ as follows:
	
\begin{eqnarray}
R_{EW} = \frac{A_{ABS} + A_{EM}}{|A_{ABS}| + A_{EM}}.
\end{eqnarray}

\begin{figure}
     \centering
%\begin{minipage}[t]{0.45\linewidth}
\includegraphics[scale=0.35]{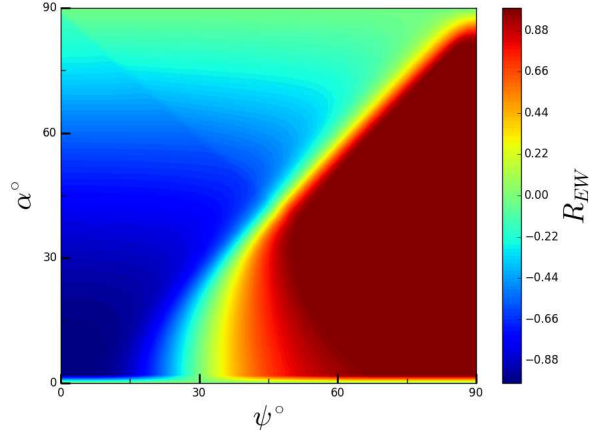}
 \caption{Map of the quantity $R_{EW} = \frac{A_{ABS}+A_{EM}}{|A_{ABS}|+A_{EM}}$,  as a function of the outflow opening angle and orientation. $R_{EW} $ quantifies the relative contribution of the emission and absorption components of the profile (see text for details).  Winds with larger values of $\psi$ are oriented away from the line of sight and become dominated by emission ($R_{EW} =1 $) for all but the largest values of $\alpha$.  In contrast, winds with smaller values of $\psi$ are aligned with the line of sight and become dominated by absorption ($R_{EW} = -1)$ for all but the largest values of $\alpha$.  Geometries approaching equivalent emission and absorption areas $(R_{EW} = 0)$ include the spherical wind, i.e., $\alpha = 90^{\circ}$, where $\alpha = \psi$, and where $\alpha = 0$ (by definition). (The remaining parameters used to generate the profiles are: $\tau_0 = 330, \ \gamma = 1, \  v_0 = 30 \ km \ s^{-1}, \ \rm{and}  \ v_{\infty}=500  \ km \ s^{-1}$.)  \\}
   \label{fig:area}
\end{figure}

\noindent
The value of $R_{EW}$ for different values of $\alpha$ and $\psi$ is shown in Figure~\ref{fig:area}. In the spherical case, while ignoring occultation from the source, one would expect the profile to have equal absorption and emission EWs, or $R_{EW} = 0$ \citep[e.g.,][]{ 2011ApJ...734...24P}.  This is not always the case for bi-conical outflows.  When viewed perpendicular to the line of sight (i.e., large values of $\psi$), a bi-conical outflow heavily favors emission, $(R_{EW} \approx 1)$, for all but the largest values of $\alpha$.  In contrast, when viewed directly along the line of sight  (i.e., small values of $\psi$), the bi-conical outflow favors absorption, $(R_{EW} \approx -1)$, for all but the largest values of $\alpha$.  When the edge of the cone is aligned with the line of sight (i.e., $\alpha < 90^{\circ}$ and $\alpha = \psi$) a large fraction of the source remains uncovered and the profile tends to have $R_{EW} \approx 0$. Small changes in either the orientation or the opening angles have large effects on $R_{EW}$ in this regime.  

          \begin{figure}
     \centering
%\begin{minipage}[t]{0.45\linewidth}
\includegraphics[scale=0.35]{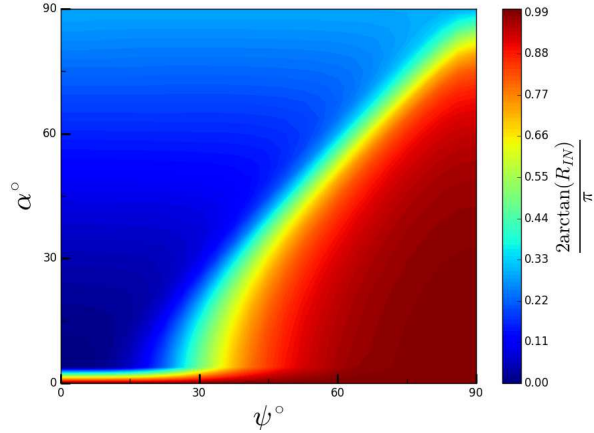}
 \caption{Map of the quantity $\frac{2\arctan{(R_{IN})}}{\pi}$, with $R_{IN} = \frac{A_{EM\backslash ABS}}{A_{ABS\backslash EM}}$, as a function of the outflow opening angle and orientation. $R_{IN}$ quantifies the contribution of the blue emission infilling to the absorption component (see text for details). Winds with larger values of $\psi$ are oriented away from the line of sight and become dominated by emission ($R_{IN} >>1 $) for all but the largest values of $\alpha$.  In contrast, winds with smaller values of $\psi$ are aligned with the line of sight and become dominated by absorption ($R_{IN} << 1)$ for all but the largest values of $\alpha$.  (The remaining parameters used to generate the profiles are: $\tau_0 = 330, \ \gamma = 1, \  v_0 = 30 \ km \ s^{-1}, \ \rm{and} \ v_{\infty}=500 \ km \ s^{-1}$.) \\}
   \label{fig:RIN}
   \end{figure}  
   
	In addition, we also investigate how the outflow geometry directly affects the blue shifted emission in-filling of the absorption component in a P~Cygni profile by allowing the geometry to vary while holding the remaining parameters constant in our models.  We isolate the emission and absorption components, and consider the ratio:

\begin{eqnarray}	
R_{IN} = \frac{A_{EM\backslash ABS}}{A_{ABS\backslash EM}},
\end{eqnarray}

\noindent where $A_{EM\backslash ABS}$ is the blue shifted emission area, excluding the absorption component, and $A_{ABS\backslash EM}$ is the absorption area, excluding the emission component.  For $R_{IN} <<1$, there is barely any blue emission in-filling the absorption component of the profile.  These outflows are characteristically oriented along the line of sight and have smaller opening angles.  For $R_{IN}>>1$, the profile is dominated by the emission component, and little to no absorption is visible in the profile.  Such outflows are oriented away from the line of sight and have smaller opening angles.  The later scenario is dramatically different from the blue emission in-filling present in spherical outflows, and unique to bi-conical outflows.  As pointed out by \citet{zhu2015near}, the blue emission in-fill can never dominate the absorption component in a spherical outflow.  Ultimately, the exact value of the ratio $R_{IN}$ will depend on all of the parameters mentioned in this paper, and additional factors such as multiple scattering, i.e., photons reemitted outside the resonant channel, but arguably the most important factor affecting $R_{IN}$ is the geometry.    

	We have plotted $(2/\pi)\arctan{(R_{IN})}$ in Figure~\ref{fig:RIN} to map the range of the ratio $R_{IN}$ between zero and one.  It is important to note that $R_{IN}$ cannot be directly observed, for the blue infill and absorption areas will be combined in a P~Cygni profile, i.e., $A_{EM\backslash ABS}$ + $A_{ABS\backslash EM}$.  Therefore, only Figure~\ref{fig:area}, not Figure~\ref{fig:RIN}, can be compared to observations.  However, the blue shifted emission infill can be separated from the true absorption component using $R_{IN}$, from Figure~\ref{fig:RIN}, if model fits can sufficiently constrain the geometry of the outflow and the observed P~Cygni profile.  In the next section, we explore how reliably the outflow parameters  can be recovered in simulated data.  %, i.e., $A_{EM\backslash ABS}$ + $A_{ABS\backslash EM}$.

\section{Simulated Spectra and Fitting procedure}

In this Section, we explore the accuracy at which the outflow parameters are recovered using spectral fitting procedures. To this aim, we created a set of 50 simulated spectra, with a range of randomly chosen input parameters. Our model is described by six parameters: the opening angle, $\alpha$, the orientation angle, $\psi$, the power law index of the velocity field, $\gamma$, the optical depth, $\tau_o$, the initial velocity, $v_0$, and the terminal velocity, $v_{\infty}$. We generated the spectra by selecting parameters uniformly from the following parameter ranges: $20^{\circ}\leq\alpha\leq90^{\circ}$, $0^{\circ}\leq\psi\leq90^{\circ}$, $.5\leq\gamma\leq4$, $.01\leq\tau_0\leq100$, $2\leq v_0 \leq 80 \ km \ s^{-1}$, and $200\leq v_{\infty} \leq 800 \ km \ s^{-1}$.  In order to reproduce actual data, we added Gaussian noise to the simulated spectra, to reach a signal to noise ratio of approximately 10 in the normalized continuum \citep{2015ApJ...809...19H}.  

%The noise was generated by shifting the mock flux values by a random number drawn from a gaussian distribution centered on zero with a standard deviation equal to the error assigned to each datum.  The errors themselves were drawn from a gaussian distribution with mean, $\mu = .13$, and standard deviation, $\sigma = .047$.  We have convolved both our model and best fits to a resolution of 25 km/s.  The error distribution and resolution were chosen from galaxy 0926-4427 presented in \citet{heckman2011extreme} who obtained spectroscopic data via the COS spectrograph.  This galaxy was chosen to provide a realistic basis for our errors.  In addition, we plan to fit our model to the galaxies (including 0926-4427)  discussed in \citet{henry2015lyalpha} in a forthcoming paper. 

To derive the best fit parameters for the model, we used the emcee package in python \citep{foreman2013emcee}, which relies on the Python implementation of Goodman's and Weare's Affine Invariant Markov Chain Monte Carlo (MCMC) Ensemble sampler \citep{goodman2010ensemble}.  
 
%We probed our parameter spaces with 30 walkers for 5,000 steps.  The initial walkers were chosen at random from within the following parameter spaces: $0 \leq \alpha  \leq90^{\circ}, 0  \leq \psi  \leq90^{\circ}, 0.01  \leq \tau_{0}  \leq 10, 2  \leq v_0  \leq 60\text{ km/s}$, and $2  \leq v_{\infty}  \leq 800\text{ km/s}$.  For $\tau_{0}$ and $v_{\infty}$ the values were selected from the log base 10 equivalent ranges.  The walkers were allowed to walk within the following boundaries: $0  \leq \alpha  \leq 90^{\circ},0  \leq \psi  \leq 90^{\circ},0.01  \leq \tau_{0}  \leq1000,2  \leq v_0  \leq 150\text{ km/s}$, and $2  \leq v_{\infty}  \leq 1500\text{ km/s}$.  The first 1000 steps, or the so-called 'burning period', were removed from each chain to improve precision. 
The emcee analysis returns probability distributions (PDFs) for each parameter. 
%The binning mechanism for the PDFs was chosen using the built in Python 'auto' binning procedure.  The procedure utilizes both the Sturgis and Freedman Diaconis bin width estimators \citep{freedman1981histogram,sturges1926choice}.  
 Because many of the PDFs are asymmetric, the best fit parameters were chosen to represent the most likely value (i.e., the mode) for a given parameter's PDF.  Similarly,  we chose the median of the absolute deviation around the mode (MAD) to describe the width for each distribution (and use this value as an estimate of the error associated with each parameter).  

\subsection{Discussion of Returned Parameters From Model Fitting}  

   \begin{figure*}[t!]
   \includegraphics[scale=0.48]{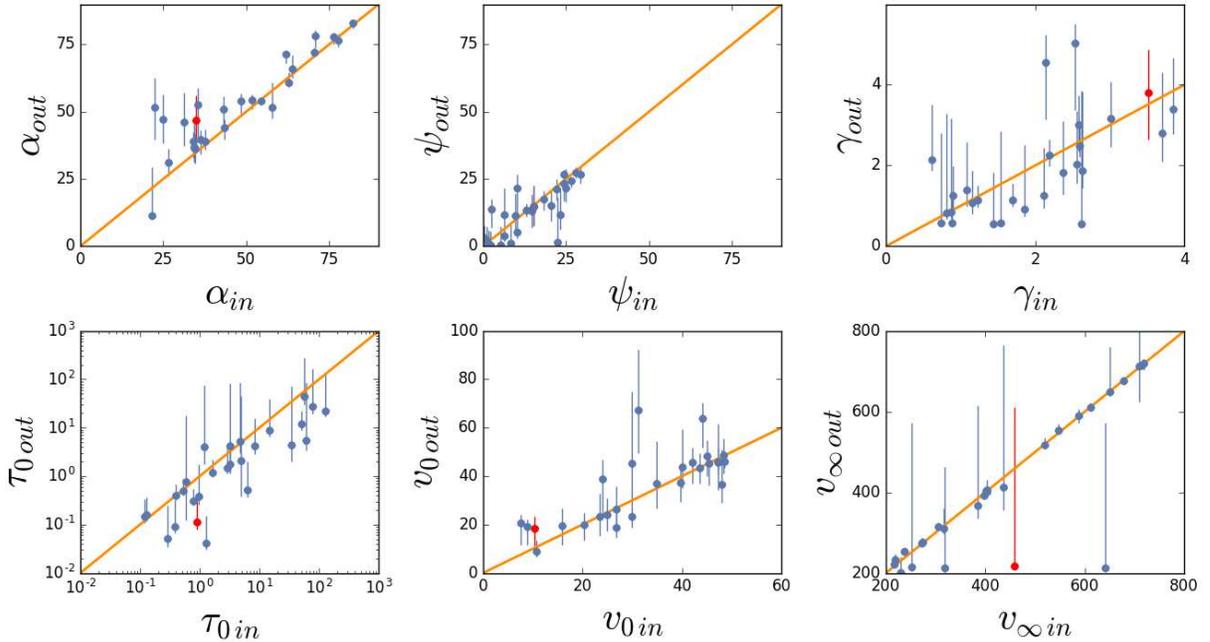}
    \caption{Results of the simulations performed to assess the accuracy at which the outflow parameters are recovered using spectral fitting procedures. This Figure shows the results for bi-conical models with orientation angle $0^{\circ}\leq \psi < 30^{\circ}$ (i.e., approximately oriented along the line of sight).  The recovered parameters are plotted as a function of the input values. The solid orange lines represent a perfect match between the true value and best fit.   For sufficiently small opening angles, $\alpha$, these profiles are characterized by large absorption and small emission components.  We have traced a profile that was drastically underestimated in $v_{\infty}$ by coloring it red.  The corresponding profile is shown in Figure~\ref{fig:small_psi}. } 
   \label{fig:30}
   \end{figure*}
   
   \begin{figure*}
   \includegraphics[scale=0.48]{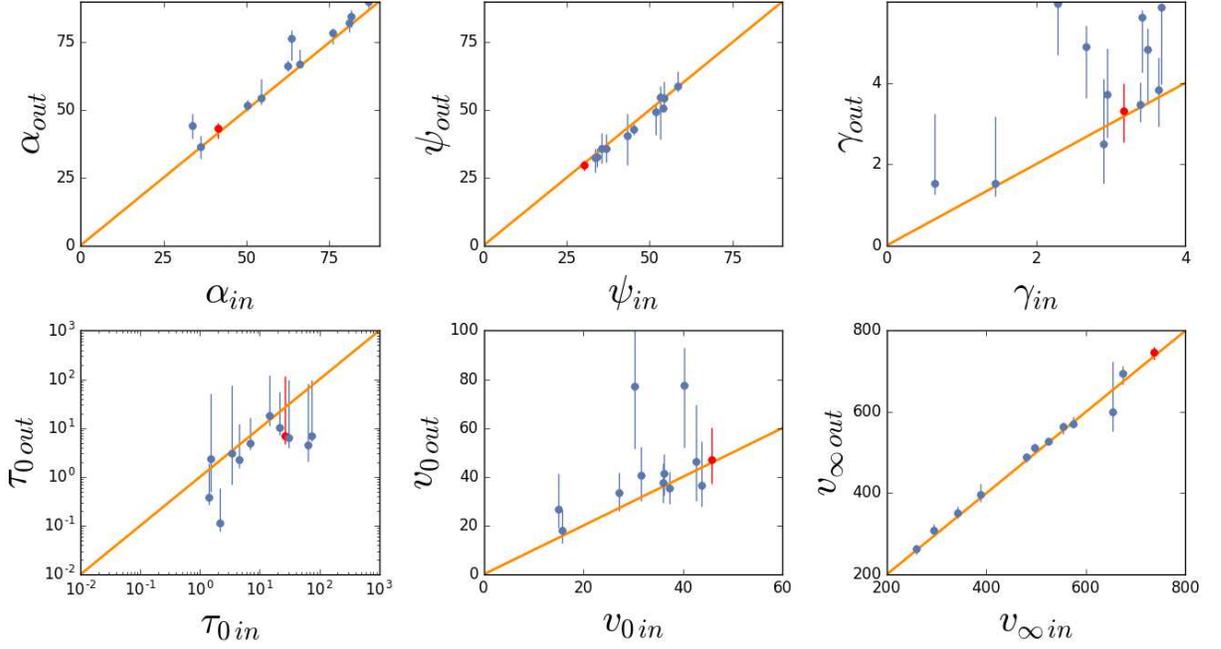}
    \caption{Same as Figure~\ref{fig:30} for  bi-conical models with orientation angle $30^{\circ}\leq \psi < 60^{\circ}$.  These profiles are characterized by having both strong absorption and emission components.  We have traced a profile for reference by coloring it red.  The corresponding profile is shown in Figure~\ref{fig:med_psi}. } 
   \label{fig:60}
      \end{figure*}
      
          \begin{figure*}
\includegraphics[scale=0.48]{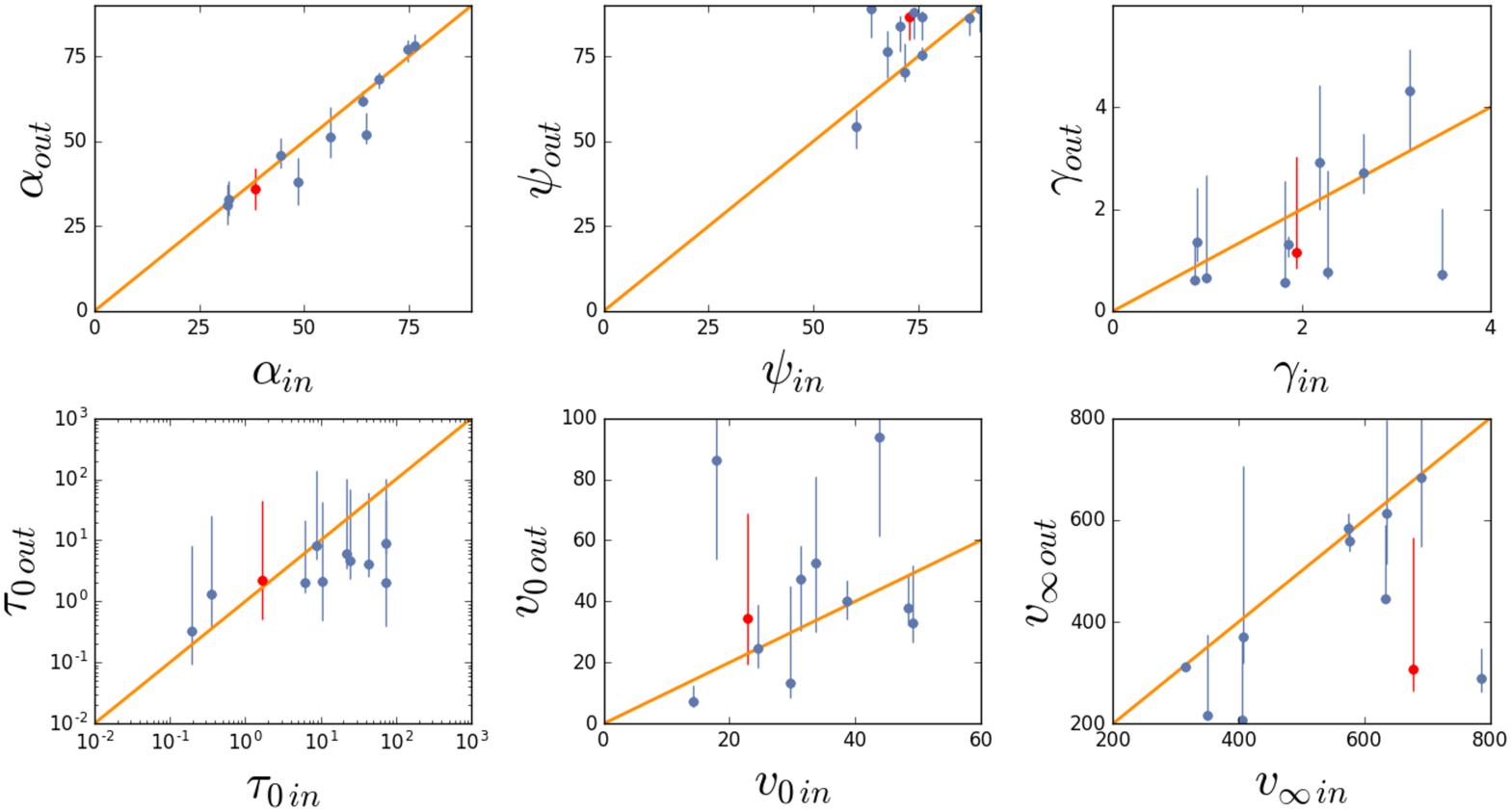}
    \caption{Same as Figure~\ref{fig:30} for  bi-conical models with orientation angle $60^{\circ}\leq \psi \leq 90^{\circ}$.  These outflows are approximately perpendicular to the line of sight.  For sufficiently small opening angles, $\alpha$, these profiles are characterized by a small (if visible) absorption dip and a large emission spike.  The terminal velocity cannot be reliably recovered from a profile with only an emission component.  We have traced one such profile in red.  The corresponding profile is shown in Figure~\ref{fig:large_psi}.} 
   \label{fig:90}
     \end{figure*}
Figures~\ref{fig:30}, \ref{fig:60}, and \ref{fig:90}  show the comparison between the input parameter values (on the horizontal axis) and the recovered values (on the vertical axis). For clarity, we have organized the simulations according to orientation.  In general, the parameters describing the geometry of the outflow ($\alpha$,$\psi$) are the best constrained. The accuracy with which we derive the velocity and optical density parameters depends  on the inclination of the cone with respect to the line of sight. The highest accuracy is reached for $30^{\circ}\leq \psi < 60^{\circ}$.  These profiles are characterized by having both a strong emission and absorption component visible in the spectra, simultaneously providing constraints on the geometry, the velocity, and the density of the scattering gas.  An example spectrum has been provided in Figure~\ref{fig:med_psi}.  Its simulation results are highlighted by the red point in Figure~\ref{fig:60}.

When the cone is oriented almost perpendicularly to the line of sight (i.e., for $\psi > 60^{\circ}$), and for small opening angles,  the profiles are dominated by the emission component. Additionally, the limited absorption present in these profiles is restricted to material close to the source of emission, with low velocity and high density.  Shells at higher velocities no longer contribute to the absorption -- limiting their diagnostic power.  In these cases, the terminal velocity $v_{\infty}$ is not well constrained (i.e., measured with large uncertainties) and it is typically underestimated.  This is because $v_{\infty}$ has little effect on the emission component.  There are two reasons for this.  The first is that shells at higher velocities (i.e. near the terminal shell) have relatively lower density compared to shells at lower velocities.  Hence, their contribution to the spectrum is weak, by comparison, to shells at higher densities.  This effect is not unique to the emission profile, but also present in how the higher velocity shells affect the absorption profile.  The second reason is due to a projection effect.  The energy absorbed by shells near the terminal velocity will be reemitted over a much larger range in observed velocity.  For example,  the energy absorbed by the blue-shifted emitting region of a spherical terminal shell will be distributed from zero to the terminal velocity.  Hence, its already small contribution to the emission profile will be diminished dramatically.  This issue is not as prominent in the absorption component because the range of observed velocities is constrained by the absorption region and does not increase with radius (or velocity) like in the emission case.  Therefore, estimates of the maximum or terminal wind velocities from emission features alone will return lower estimates of these values when compared to studies including the absorption features. 

An example spectrum demonstrating this effect has been provided in Figure~\ref{fig:large_psi}.  Its simulation results are highlighted by the red point in Figure~\ref{fig:90}.  Notice that the recovered best fit nearly matches the original spectrum, or true value, however, the terminal velocity has been severely underestimated signifying the negligible contribution the shells near terminal velocity make to the spectrum.  It is important to realize that the inability for the emission component alone to constrain the terminal velocity is independent of geometry.

 $\psi$ itself appears to be overestimated in Figure~\ref{fig:90}.  As demonstrated in Section $4$, $\psi$ acts to control the position of the emission component and a large change in $\psi$ results in a small shift in the actual profile in terms of the observed velocities.  This minuscule effect is difficult to detect with low signal-to-noise data.

Finally, when the cone is parallel to the line of sight, the parameters are typically well recovered, albeit with a larger uncertainty. We have selected one case, with drastically underestimated $v_{\infty}$.  The profile is plotted in Figure~\ref{fig:small_psi} and the simulation results are highlighted by the red point in Figure~\ref{fig:30}. It is easy to see that, because of the low column density of this particular example, the true velocity range of the absorption component has been lost due to the low signal-to-noise, resulting in an underestimate of $v_{\infty}$.

\begin{center}
\includegraphics[scale=.45]{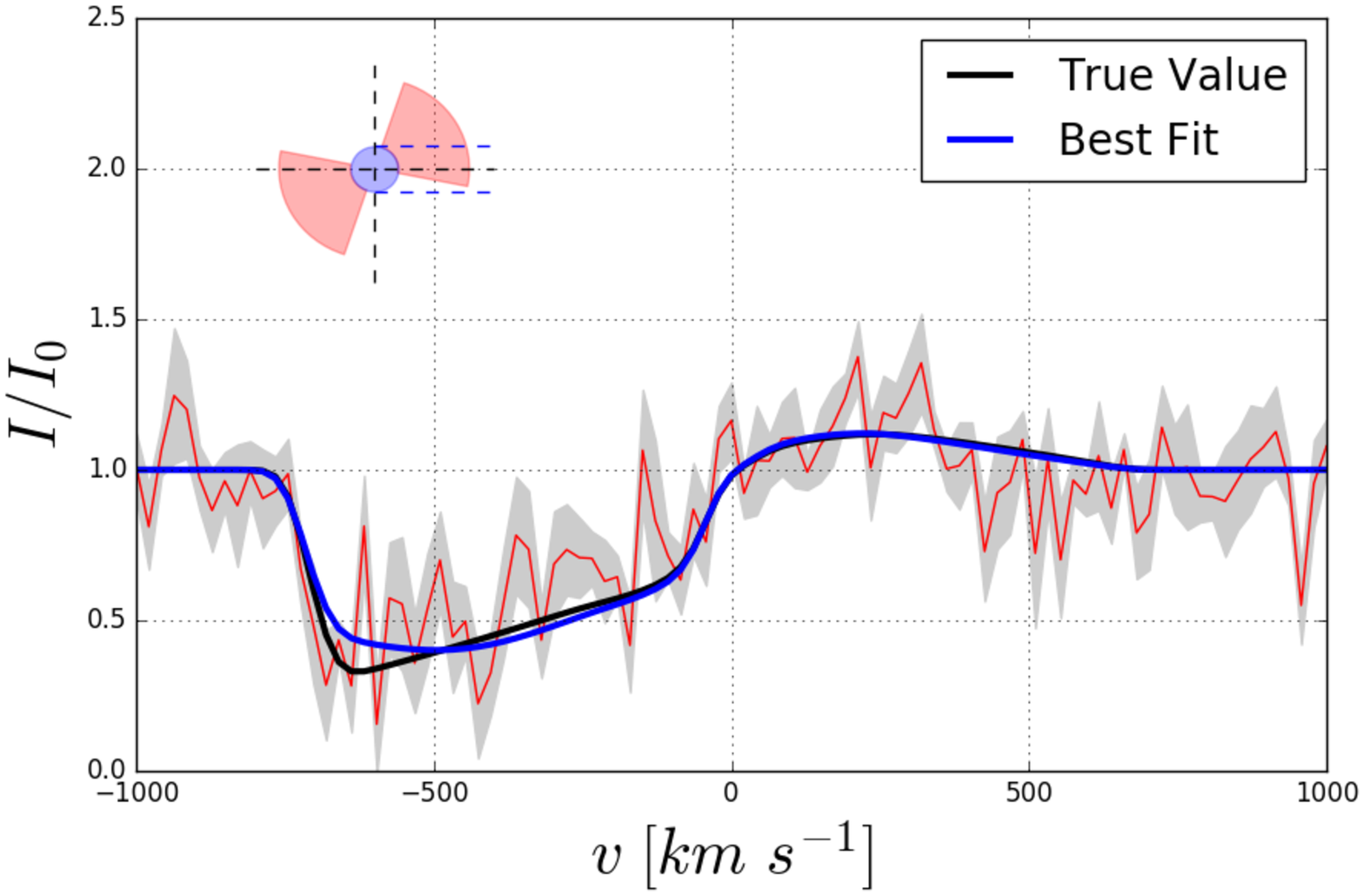}
\captionof{figure}{Mock profile (red), true value (black), and best fit (blue) for the red colored marker in Figure~\ref{fig:60}.  The grey area represents plus or minus the error at each observed velocity.  Here, the outflow is oriented such that the profile will have both a strong absorption and emission component ($30 \leq \psi < 60^{\circ}$).  An emblem representing the outflow geometry as viewed from the right has been placed in the upper left corner.  Parameters are most easily returned for these types of profiles.}
 \label{fig:med_psi}
\end{center}
\begin{center}
\includegraphics[scale=.45]{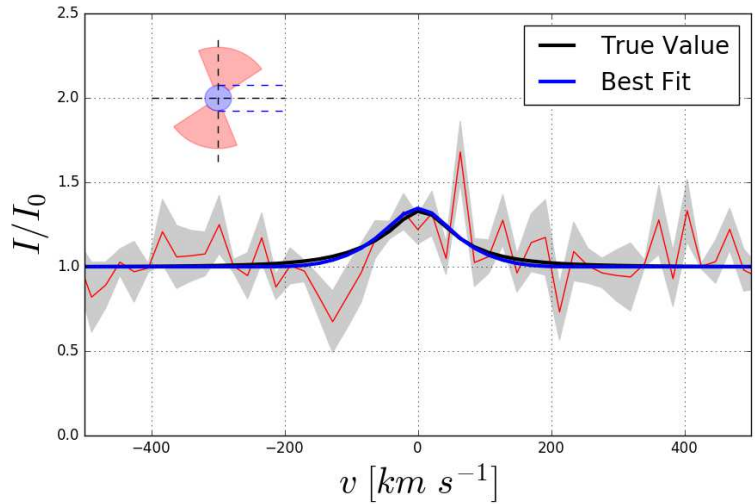}
\captionof{figure}{Same presentation as in Figure~\ref{fig:med_psi}, but for the red colored marker in Figure~\ref{fig:90}.  Here, the outflow is oriented perpendicular to the line-of-sight ($60^{\circ}\leq \psi$). Hence, the profile is dominated by an emission spike with no visible absorption.  Shells near the terminal velocity make negligible contributions to the emission profile rendering the terminal velocity difficult to recover.  Indeed, the best fit almost matches the true value, however, the terminal velocity has been severely underestimated.  }
 \label{fig:large_psi}
\end{center}
\begin{center}
\includegraphics[scale=.45]{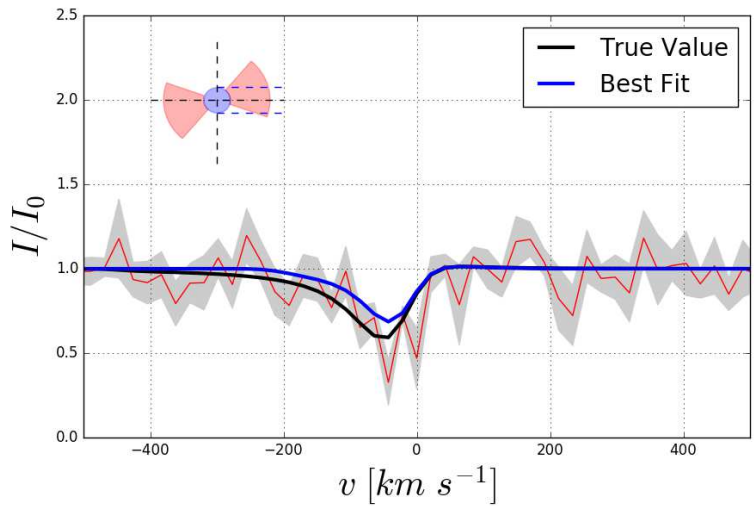}
\captionof{figure}{Same presentation as in Figure~\ref{fig:med_psi}, but for the red colored marker in Figure~\ref{fig:30}.  Here, the outflow is oriented toward the observer ($\psi < 30^{\circ}$), and confined to the absorption region.  Hence, the profile is dominated by an absorption dip with little to no visible emission.  It is difficult for the model to find the correct terminal velocity, for the maximum velocity at which absorption occurs can easily succumb  to the signal to noise. }
 \label{fig:small_psi}
\end{center}

\section{Conclusions}
In this paper we have presented and discussed an extension to the SALT model presented in \citet{Scarlata:2015fea}, to calculate the expected absorption and
re--emission line profile generated in a bi-conical outflow surrounding a spherical source, or galaxy, with a finite size ($R_{SF}$), under the Sobolev approximation. We parametrize the outflow with the opening angle $\alpha$ and the inclination with respect to the line of sight.  Similar to our previous work, we computed the analytical profiles for a gas velocity increasing with distance from the galaxy ($v\propto r^{\gamma}$), and a constant mass outflow rate ($n_l(r)\propto (vr^2)^{-1}$).  The particular velocity and density laws used in this work are motivated by the need to explore how the absorption/emission line profiles change going from a spherical outflow (studied in Scarlata \& Panagia 2015) to a bi-conical geometry, with the same $v$ and $n$ laws.  Similarly, for simplicity in this paper we have assumed a constant ionization state of the gas throughout the outflow.  A full analysis that solve for the ionization state as a function of the density, velocity field, luminosity and spectral shape of the source is beyond the scope of this work, and will be the subject of future papers. 

We analyzed the effects of the wind geometry on the line profiles, and compared to the simpler spherical model, with constant covering fraction, $f_c$. 
The resulting profiles vary substantially. Depending on the orientation,  the profile can vary from pure absorption (i.e., small $\alpha$ and $\psi$) to pure emission (for large $\psi$ and small $\alpha$), with a range of P~Cygni-like profile shapes in between. We have studied how the ratio between the EW of the emission and absorption components changes with the outflow geometry. 

We used simulated spectra to study the accuracy and degeneracies in recovery of the outflow parameters from fitting models to the data. We show that for a typical S/N ratio of $\simeq$ 10 in the continuum, the geometry can be accurately reproduced. The velocity field (described by $v_0$, $v_{\infty}$, and $\gamma$) is better constrained when the terminal shell of the outflow intercepts the line of sight, producing absorption at $v_{\infty}$.  If this is not the case, then the maximum velocity remains largely unconstrained by the resulting absorption profiles. This fact has an interesting consequence. Because the outflows will be distributed randomly in the sky, one would expect that for many of them gas moving at $v_{\infty}$ will not necessarily contribute to the absorption component. Consequently, one would expect to find only weak correlations between the maximum velocity of absorption and galactic properties, as found by \citet[e.g.,][]{2012ApJ...760..127M}.  Furthermore, studies analyzing only emission spectra cannot reliably recover the terminal velocity in high velocity winds and are likely to return a lower estimate of the terminal velocity in comparison to studies including the absorption component.  This is in regard to any geometry of the outflow for the shells with higher velocities make negligible contributions to the emission spectrum.          

To conclude, by simultaneously modeling the resonant absorption and the associated resonant emission, the line profiles computed with
our SALT model can be used to constrain the real 3-dimensional geometry and orientation of gaseous outflows, their density field and
the velocity structure within the winds. This information can be used to constrain the wind launching mechanism, to ultimately shed light on SF driven feedback.




\appendix

\subsection{Multiple Scattering}

Here we review the method of multiple scattering by \citet{Scarlata:2015fea}.  Once a resonant photon is scattered/absorbed it will have a probability of getting re-absorbed/re-scattered by local ions.  We define the escape probability $\beta$ of a photon from a shell of velocity $v$ to be the probability that a photon is scattered/absorbed at resonance per mean free path, i.e., 
\begin{eqnarray}
\beta = \frac{1-e^{-\tau(v)}}{\tau}.
\end{eqnarray}

\noindent
Once a photon is scattered, it will have a probability $p_F$ of being reemitted via the fluorescent channel and a probability $p_R$ of being reemitted via the resonant channel.  Because of the Sobolev approximation, a photon emitted via the fluorescent channel will escape the outflow and make its way to the observer.  Of the fraction $p_R$, a fraction $1-\beta$ of photons will be absorbed again before they escape the shell.  Accordingly, some of these photons will enter the resonant channel and some the fluorescent channel.  The process repeats itself iteratively.  The final fraction of photons entering the fluorescent channel becomes 
\begin{eqnarray}\label{1}
F_F(\tau) = p_F/[1-p_R(1-\beta)],
\end{eqnarray}
\noindent
and the final fraction entering the resonant channel becomes
\begin{eqnarray}\label{2}
F_R(\tau) = \beta p_R/[1-p_R(1-\beta)].
\end{eqnarray}

	After each scattering, the optical depth will depend upon the velocity and the new trajectory of the scattered photon.  To account for this, we take $\beta$ in expressions \ref{1} and \ref{2}  averaged over the 2-sphere,
\begin{eqnarray}
\int \beta d \Omega = \frac{1}{4\pi}\int \beta d(\cos \theta) d\phi.
\end{eqnarray} 

\newpage

\subsection{Observing Aperture}

\begin{center}
\includegraphics[scale=.6]{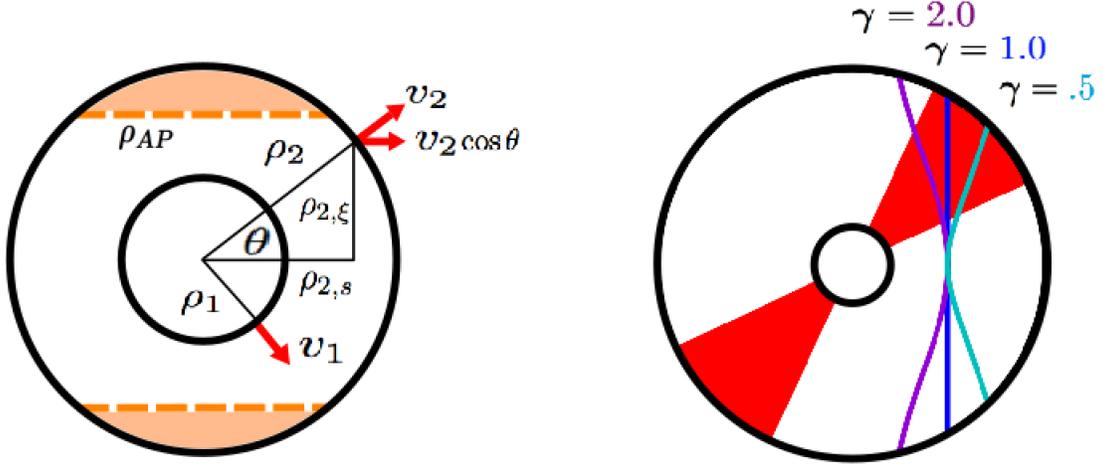}
\captionof{figure}{On the left are concentric shells of radii $\rho_1$ and $\rho_2$ at velocities $v_1$ and $v_2$, respectively.  The orange regions are excluded by an aperture of radius $\rho_{AP}$.  For a shell of velocity $v_1$, the equivalent observed velocity for a shell with larger radius, $\rho_2$, is defined by the expression, $v_1 = v_2 \cos{\theta}$.  We want to find this location in terms of position, hence, we use the angle $\theta$ to construct a right triangle in terms of the radius, $\rho_2$, i.e., $\rho_{2}^2 = \rho_{2,s}^2+\rho_{2,\xi}^2$, where $\rho_{2,s}=\rho_2\cos{\theta}$ and $\rho_{2,\xi}=\rho_2\sin{\theta}$ are the horizontal and vertical components of the triangle, respectively.  A shell will be excluded by the aperture if the location of constant observed velocity lies outside the aperture radius, i.e. $\rho_{2,\xi}>\rho_{AP}$.  To the right are curves, $\Gamma$, of constant observed velocity, $v_{obs}$, for velocity fields of different power law index, $\gamma
$, drawn overtop a bi-conical outflow shown in red.  These curves represent where the surfaces of constant observed velocity intersect the plane of the diagram.  If a shell intersects the curve, it will contribute to the spectrum at the observed velocity defining $\Gamma$.}
 \label{fig:concentric_shells}
\end{center}

%\begin{center}
%\includegraphics[scale=.55]{path_of_cv.png}
%\captionof{figure}{Paths of constant observed velocity for power law $\gamma $.  The paths are drawn for a shell of velocity $v$.  Only shells that lie along the path will contribute to the observed velocity at $v$. }
% \label{fig:constant_vel}
%\end{center} 

To account for a limiting, observing aperture we need to make precise when a shell in an outflow will no longer contribute to the observed spectrum.  For a given $v_{obs}$, we define the curve of constant observed velocity, $\Gamma$, to be the intersection of the surface of constant $v_{obs}$ with the $\xi s$-plane.  We have plotted $\Gamma$ for several velocity fields of different power law index $\gamma$ to the right in Figure \ref{fig:concentric_shells}.  Shells that intersect $\Gamma$ outside of the aperture radius, $\rho_{AP}$, will no longer contribute to the observed spectrum at $v_{obs}$.     

We seek to parameterize $\Gamma$ in the $\xi s$-plane.  To this end, we consider the generic problem of identifying the curve of constant observed velocity between two shells of known position in a velocity field of arbitrary power law as defined in Section 2.  Let $\rho_1,v_1$ and $\rho_2,v_2$ be the radii and velocities, respectively, for two separate shells such that $\rho_2 > \rho_1$.  See Figure \ref{fig:concentric_shells}.  Finding the equivalent observed velocity to $v_1$ at $\rho_2$, we get

\begin{eqnarray}
v_1 &=& v_2\cos{\theta}\\
\implies \cos{\theta} &=& \frac{v_1}{v_2} = \left(\frac{\rho_1}{\rho_2}\right)^{\gamma}.
\end{eqnarray}

We want to exclude all shells along the curve of constant observed velocity that fall outside the aperture radius, $\rho_{AP}$, hence, we seek the location of constant observed velocity in terms of position.  Using the angle determined from the velocity field, we define a triangle in terms of the radius, $\rho_2$, such that $\rho_{2,s}^2+\rho_{2,\xi}^2 = \rho_2^2$, where $\rho_{2,s} = \rho_2\cos\theta$ and $\rho_{2,\xi}=\rho_2\sin\theta$ are the horizontal and vertical components of the triangle, respectively.  Moreover,  

\begin{eqnarray}
 \frac{\rho_{2,s}}{\rho_2} &=& \left(\frac{\rho_1}{\rho_2}\right)^{\gamma}\\
 \implies \rho_{2,s} &=& \rho_1^{\gamma}\rho_2^{1-\gamma}.
 \end{eqnarray}
 \noindent
 Hence, the vertical component becomes
 
 \begin{eqnarray}
 \rho_{2,\xi} = [\rho_2^2-\rho_1^{2\gamma}\rho_2^{2(1-\gamma)}]^{1/2}.
 \end{eqnarray}
 \noindent Thus, the parameterization of $\Gamma$ in the $\xi s$-plane becomes 
 \begin{eqnarray}
 \Gamma = (\xi,s) &=&  (\rho_1^{\gamma}\rho_2^{1-\gamma}, [\rho_2^2-\rho_1^{2\gamma}\rho_2^{2(1-\gamma)}]^{1/2}) \\
 &=& (xy^{(1-\gamma)/\gamma},[y^{2/\gamma}-x^2y^{2(1-\gamma)/\gamma}]^{1/2}),
 \end{eqnarray}
 
\noindent where in the last expression we have converted back into velocity and expressed the generic quantities in terms relevant to the paper.  Thus, following along the curve of constant observed velocity for a shell of radius $\rho_1$ and velocity $v_1$, all contributions to the spectrum at the observed velocity $v_1$ from shells of larger radius, $\rho_2$, will be excluded if $\rho_{AP} < \rho_{2,\xi}$, i.e., 

\begin{eqnarray}
\rho_{AP} <  [\rho_2^2-\rho_1^{2\gamma}\rho_2^{2(1-\gamma)}]^{1/2}.
\end{eqnarray}

\noindent
Converting to velocity and the relevant quantities, we get the following condition,

\begin{eqnarray}
y_{ap} < [y^{2/\gamma}-x^2y^{2(1-\gamma)/\gamma}]^{\gamma/2},
\end{eqnarray}
\noindent
where $y_{ap}  = v_{ap}/v_0$.  Therefore, we define the aperture factor

\begin{eqnarray}
\Theta_{AP} &:=& \Theta(y_{\text{ap}} - [y^{2/\gamma}-x^2y^{2(1-\gamma)/\gamma}]^{\gamma/2} ),
\end{eqnarray}

\noindent
where $\Theta$ is the Heaviside function:

\begin{gather} 
\Theta := 
\begin{cases}
0 & \rm{if \ }  \ y_{\text{ap}}  < [y^{2/\gamma}-x^2y^{2(1-\gamma)/\gamma}]^{\gamma/2} \\[1em]
1 & \rm otherwise.\\
\end{cases}
\end{gather}

\noindent
This scaling factor will exclude all shells outside the aperture radius.  Note that this construction will work for all outflow geometries.    

\newpage

\subsection{Geometric Factor}
\begin{figure*}[b!th]
  \centering
\includegraphics[scale=0.55]{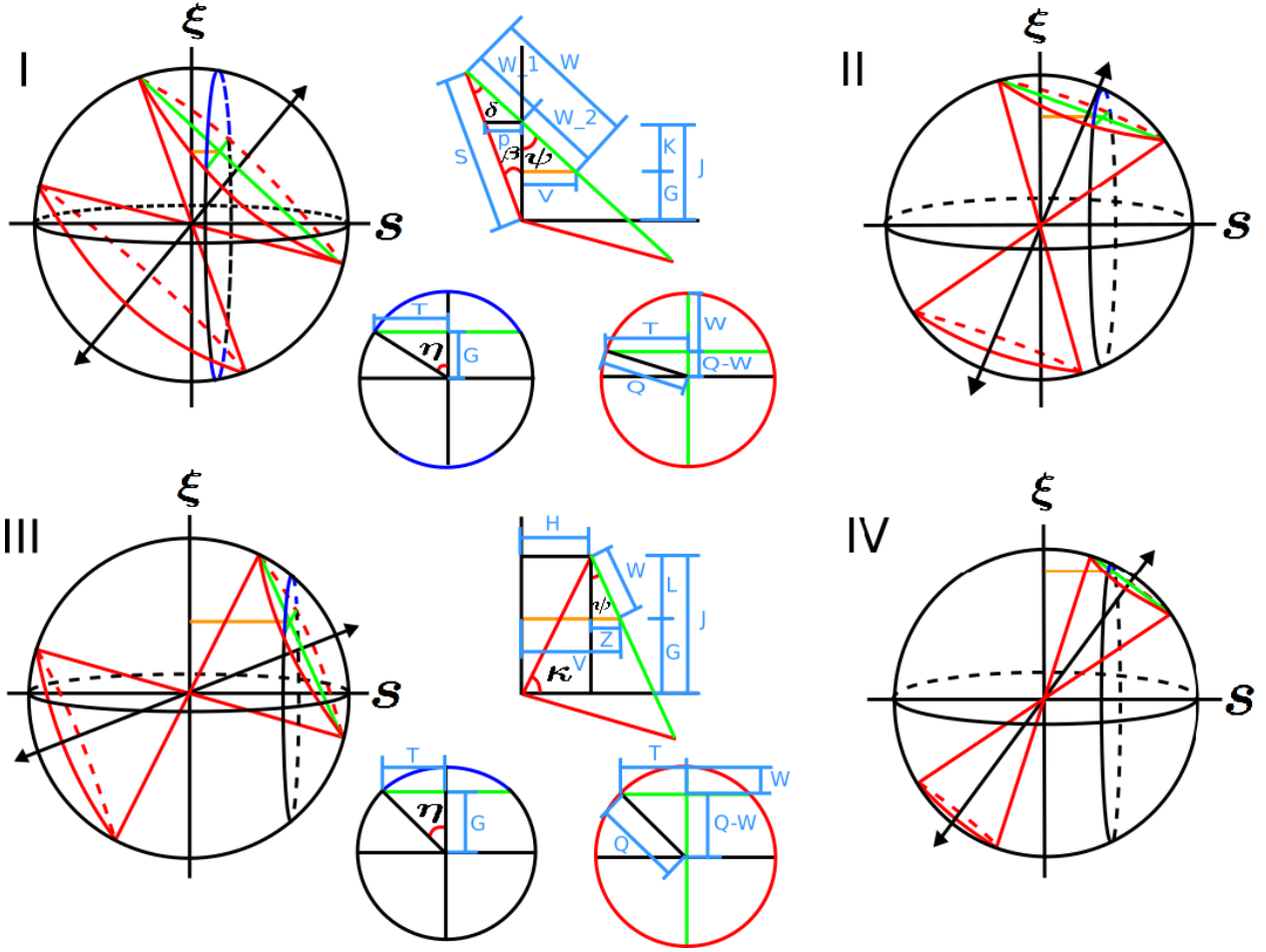}
 \caption{Diagram containing various geometries and orientations of the bi-conical outflow:  (I) $\alpha + \psi > 90^{\circ}$ and $\alpha > \psi$, (II) $\alpha + \psi > 90^{\circ}$ and $\alpha < \psi$, (III) $\alpha + \psi < 90^{\circ}$ and $\alpha > \psi$, and (IV) $\alpha + \psi < 90^{\circ}$ and $\alpha < \psi$.  We want to find the fraction of the bi-conical shell overlapping the spherical shell for a given observed velocity.   We use the band contour of constant observed velocity for a shell of velocity $y$ located at a distance $V = xy^{(1-\gamma)/\gamma}$ from the $\xi$-axis.  Our goal to compute $f_g$, is to calculate the angular separation, $\eta$, made by half of the arc length of the overlap of the spherical shell with the observed velocity contour, shown in dark blue.  To do this, we calculate the two lengths, $T$ and $G$, forming the tangent of $\eta$.  To complete this task, we calculate $T$ from the circle formed by the intersection of the bi-conical shell with the spherical shell, shown in red.  We have shown the necessary values for calculating $\eta$ in I and II.  The computation for III, is the same as I, and the computation for IV, follows from III.     }
   \label{fig:psicrosssec}
\end{figure*}

To calculate the geometric factor, $f_g$, we want to find the fraction of a bi-conical shell of intrinsic velocity, $y$, overlapping the constant observed velocity contour of interest from the corresponding spherical shell.  For a velocity field of power law index, $\gamma$, the curve of constant observed velocity, $x$, will be
\begin{eqnarray*}
\Gamma = (\xi,s) = (xy^{(1-\gamma)/\gamma},[y^{2/\gamma}-x^2y^{2(1-\gamma)/\gamma}]^{1/2}).
\end{eqnarray*}
See Appendix 1. for a derivation.  All shells that intersect $\Gamma$ will contribute to the spectrum at $x$.  Thus for a given $x$, we want to compute $f_g$ for all shells of intrinsic velocity $y$ that intersect $\Gamma$.   Then given a shell of intrinsic velocity, $y$, the distance to the relevant contour from the center of the outflow, written in terms of velocity, will be $V = xy^{(1-\gamma)/\gamma}$.  This distance has been drawn in orange in Figure \ref{fig:psicrosssec}.  

Now that we know the precise location of the observed velocity contour in terms of $x$, we need to explicitly compute $f_g$.  Our approach will be to find the angular separation, $\eta$, made by half of the arc length of the overlap of the spherical shell with the observed velocity contour, shown in dark blue.  We consider four major cases regarding the wind geometry.

For cases I and II: $\alpha + \psi > 90^{\circ}$ and $\alpha > \psi$ (I) or $\alpha < \psi$ (II).  As shown in I, the observed velocity contour touches both the front and backside of the bi-conical shell.  We will distinguish these contributions as $f_{gu}$ and $f_{gl}$, respectively.  We provide only the pictorial representation of the geometry necessary to calculate $f_{gu}$, however, calculations for both $f_{gu}$ and $f_{gl}$ are provided.  We will need the following quantities to compute $f_{gu}$:

\begin{eqnarray*}
\delta &=& \pi/2-\alpha; \ \beta = \alpha+\psi - \pi/2;\\
W_1 &=& \frac{y\sin{(\beta)}}{\sin{(\pi-\psi)}}; \ W_2 = \frac{V}{\sin{(\psi)}}; \ W = W_1+W_2;\\
P &=& \frac{W\sin{(\delta)}}{\sin{(\alpha+\psi)}}; \ S = y^{1/\gamma}; \ Q = S\sin{(\alpha)}; \ T = (Q^2-[Q-W]^2)^{1/2};\\
K &=& \frac{V}{\tan{(\psi)}}; \ J = \frac{P}{\tan{(\beta)}}; \ G = J - K; \ \eta = \arctan{\left(\frac{T}{G}\right)}.
\end{eqnarray*}
Thus, 
\begin{eqnarray*}
f_{gu} = \frac{\eta}{\pi}
\end{eqnarray*}
(If $V>J\tan{(\psi)}$, set $G = K - J$ and take $f_{gu} = 1 - \frac{\eta}{\pi}$.)  

To compute $f_{gl}$, we will need the following:
\begin{eqnarray*}
N &=& \frac{V}{\sin{\psi}}; \ L = W_1 - N; \ T = [Q^2-(Q-L)^2]^{1/2};\\
H &=& \frac{V}{\tan{\psi}}; \ G = J + H; \ \eta = \arctan{\left(\frac{T}{G}\right)}.
\end{eqnarray*}
Thus,
\begin{eqnarray*}
f_{gl} = \frac{\eta}{\pi}.
\end{eqnarray*}
Therefore,
\begin{eqnarray*}
f_{g} = f_{gu}+f_{gl}.
\end{eqnarray*}

For cases III and IV: $\alpha + \psi < 90^{\circ}$ and $\alpha > \psi$ (III) or $\alpha < \psi$ (IV).  We will need the following quantities to compute $f_g$:
\begin{eqnarray*}
\kappa &=& \alpha + \psi; \ H = y^{1/\gamma}\cos{(\kappa)}; \ Z = V - H; \ W = \frac{Z}{\sin{(\psi)}}; \ L = \frac{Z}{\tan{(\psi)}};\\
 J &=& y^{1/\gamma}\sin{(\kappa)}; \ Q = S\sin{(\alpha)}; \ G = J - L; \ T = [Q^2-(Q-W)^2]^{1/2}; \\
  \eta &=& \arctan{\left(\frac{T}{G}\right)}.
\end{eqnarray*}
Therefore,
\begin{eqnarray*}
f_g = \frac{\eta}{\pi}.
\end{eqnarray*}
(If $V>J\tan{(\psi)}+H$, set $G = L-J$ and take $f_g = 1 - \frac{\eta}{\pi}$.)



\end{document}